\documentclass{article}
\usepackage{amssymb,amsfonts,amsmath,hyperref}

\newcommand{\<}{\langle}
\newcommand{\R}{\mathcal{R}}
\newcommand{\C}{\mathcal{C}}
\newcommand{\E}{\mathcal{E}}

\newcommand{\K}{\mathcal{K}}
\newcommand{\D}{\mathcal{D}}

\newcommand{\N}{\mathcal{N}}
\newcommand{\HS}{{\scriptscriptstyle\mathrm{HS}}}

\def\p{\partial}

\newcommand{\mr}{\mathrm}

\newcommand{\mR}{\mathbb{R}}
\newcommand{\mN}{\mathbb{N}}
\newcommand{\mZ}{\mathbb{Z}}

\newcommand{\dsp}{\displaystyle}

\newcommand{\then}{\Rightarrow}

\newcommand{\ep}{\epsilon}
\newcommand{\vep}{\varepsilon}

\newcommand{\vp}{\varphi}

\newcommand{\con}{\mathrm{const}}

\newcommand{\lp}{\left(}
\newcommand{\rp}{\right)}

\newcommand{\la}{\lambda}
\newcommand{\La}{\Lambda}
\newcommand{\ov}{\overline}

\newcommand{\hb}{h_{za}^B}
\newcommand{\wh}{\widehat}
\newcommand{\hp}{\widehat{\psi}}
\newcommand{\da}{\delta}

\DeclareMathOperator{\id}{id}

\DeclareMathOperator*{\Car}{\times}

\DeclareMathOperator{\Tr}{Tr}

\title{Quantum backreaction (Casimir) effect\\II. Scalar and electromagnetic fields}
\author{Andrzej Herdegen%
\thanks{e-mail: herdegen@th.if.uj.edu.pl}\\
{\it Institute of Physics, Jagiellonian University,}\\
{\it Reymonta 4, 30-059 Cracow, Poland}}
\date{}
\begin{document}

\maketitle

\begin{abstract}
Casimir effect in most general terms may be understood as a
backreaction of a quantum system causing an adiabatic change of
the external conditions under which it is placed. This paper is
the second installment of a work scrutinizing this effect with the
use of algebraic methods in quantum theory. The general scheme
worked out in the first part is applied here to the discussion of
particular models. We consider models of the quantum scalar field
subject to external interaction with ``softened'' Dirichlet or
Neumann boundary conditions on two parallel planes. We show that
the case of electromagnetic field with softened perfect conductor
conditions on the planes may be reduced to the other two. The
``softening'' is implemented on the level of the dynamics, and is
not imposed \emph{ad hoc}, as is usual in most treatments, on the
level of observables. We calculate formulas for the backreaction
energy in these models. We find that the common belief that for
electromagnetic field the backreaction force tends to the strict
Casimir formula in the limit of ``removed cutoff'' is not
confirmed by our strict analysis. The formula is model dependent
and the Casimir value is merely a term in the asymptotic expansion
of the formula in inverse powers of the distance of the planes.
Typical behaviour of the energy for large separation of the plates
in the class of models considered is a quadratic fall-of.
Depending on the details of the ``softening'' of the boundary
conditions the backreaction force may become repulsive for large
separations.

\end{abstract}
\vfill \eject

\sloppy
\renewcommand{\theequation}{\thesection.\arabic{equation}}

\section{Introduction}\label{intII}

This is the second of the two papers devoted to the Casimir effect
in which we develop more fully what was announced in \cite{her}.
In these papers we advocate the use of the algebraic approach to
the quantum systems as the most natural setting for the discussion
of the Casimir effect.  This approach gives a clear understanding
of the sources of the difficulties one encounters in more
traditional treatments, and allows a mathematically rigorous
analysis of the effect. In the first paper \cite{AH} this analysis
was carried out on a more general level, for a wide class of
quantum systems and external conditions. We derived a general
criterion for admissible models and obtained formulae for the
backreaction energy and generalized force. The reader should refer
to \cite{AH} for the background and our statement of the problem,
and the results mentioned here. Sections 1 -- 5 of that paper form
prerequisites for the present one, and in what follows we assume
their knowledge by the reader. Also, we refer the reader to
\cite{AH} for bibliography.

The main results and outline of the paper are as follows. In
Section \ref{sc} we show how the quantum scalar field with
external interaction of planar symmetry fits into the scheme
discussed in Sections I-3 and I-5. More precisely, the perturbing
interaction is assumed to modify the $z$-motion only (the system
is then translationally symmetric in the directions orthogonal to
the $z$-axis). The necessary preliminary condition (ii) of Sec.\
I-5 for the admissibility of the model is formulated in terms of
the $z$-motion generator. In Section \ref{eq} we obtain the
necessary and sufficient condition for the equivalence of the
vacuum representation and the ground state representation of the
field influenced by the perturbation. In Section \ref{en} this
condition is then extended to be also necessary and sufficient for
the existence of finite backreaction energy. Next, in Section
\ref{mdn} we propose a class of models of the field perturbation
imitating the Dirichlet or Neumann conditions on two parallel
planes separated by an adiabatically changeable distance. The
models depend on two functions whose role is to soften the effect
of the planes for high energies of the particles. For a class of
these functions the admissibility conditions are satisfied and an
explicit formula for the Casimir energy is obtained. We
investigate this formula in Section \ref{asy}. We show that the
original Casimir expression constitutes a~term in the expansion of
the formula in inverse powers of the distance of the planes. Using
a scaling property of the formula we can approximate the strict
boundary conditions. In the case of Dirichlet conditions the
Casimir energy becomes infinite, but the Casimir force may be
interpreted to approach the Casimir value in the limit (although
with qualifications). However, in the Neumann case both energy and
force become meaningless in the limit. We view this as a typical
situation, the Dirichlet case being exceptional. Section \ref{el}
treats the electromagnetic case with metallic boundary conditions.
We show that this model may be reduced to the superposition of
scalar Dirichlet and Neumann cases. Therefore, there is no strict
boundary limit for this model. We find that in the class of models
considered here typical fall-off of the backreaction force for
large separation of the plates is by one order weaker than in the
original Casimir formula, and the force may become repulsive in
the limit. Appendices contain more technical results needed in the
main text.

Precise formulation of the central results of the paper is given
in (Mod) in Section~\ref{mdn}, (Asym) in Section \ref{asy}, and in
Section \ref{el}. Also, more extensive discussion of their
physical significance will be found in the opening parts of
Sections \ref{mdn} and~\ref{asy}, and in the closing part of
Section \ref{el}.

\setcounter{equation}{0}

\section{Scalar field under external conditions with planar symmetry}\label{sc}

We apply here the formalism of Sections I-3 and I-5 to more
specific models. We take the real Hilbert space $\R$ of Sec.\ I-3
to be the tensor product of a space $\R_\bot$ of the motion in the
$x$-$y$ plane, and of a space $\R_z$ of the motion in the
$z$-direction, $\R=\R_\bot\otimes\R_z$. For the complexified
versions of these spaces we also have $\K=\K_\bot\otimes\K_z$. Let
a positive operator $h_\bot$ in $\K_\bot$, with domain
$\D(h_\bot)$, describe the perpendicular motion, and a positive
operator $h_z$ in $\K_z$, with domain $\D(h_z)$, describe the
$z$-motion. (Both operators are assumed to commute with the
complex conjugation.) Then the operator $h$ in $\K$ defined in
standard way by the form method as
\begin{equation}\label{sc-h}
 \D(h)=\D(h_\bot\otimes\id)\cap\D(\id\otimes h_z)\,,\quad
 h=\sqrt{(h_\bot\otimes\id)^2+(\id\otimes h_z)^2}\,,
\end{equation}
is a densely defined, selfadjoint, positive operator. More
precisely, the prescription
\begin{equation}\label{sc-f}
 q(\psi,\vp)=\big((h_\bot\otimes\id)\psi,(h_\bot\otimes\id)\vp\big) +
 \big((\id\otimes h_z)\psi,(\id\otimes h_z)\vp\big)
\end{equation}
defines a closed quadratic form on
$\D(h_\bot\otimes\id)\cap\D(\id\otimes h_z)$. This form determines
in the standard way the unique positive operator $h$ by
$q(\psi,\vp)=(h\psi,h\vp)$. This operator defines a quantum model,
as described in Sec.\ I-3.

The free quantum scalar field model fits into this scheme with the
choices $\K_\bot=L^2(\mR^2,dx\,dy)$, $\K_z=L^2(\mR,dz)$,
$h^2_\bot=-\Delta_\bot$, $h^2_z=-\p_z^2$, where $\Delta_\bot$ is
the two-dimensional Laplacian. We now want to introduce external
conditions which modify the $z$-motion, while leaving the
transversal motion unchanged. However, if we leave the strict
translational symmetry in the $x$-$y$ plane intact, then the bound
(I-5.8), which gives the condition for the implementability of the
modified dynamics in the original representation (technically: the
implementability of an appropriate Bogoliubov transformation),
cannot be satisfied. Both mathematical and physical reason is
clear: this condition states that in the new state the particle
number $\N_a$ is finite, which cannot be satisfied due to the
translational symmetry in the $x$-$y$ plane. Nevertheless, we can
apply the standard ``thermodynamic limit'' procedure for the
transversal directions: we restrict the \mbox{$x$-$y$} motion to
some compact region, demand finiteness of particle number $\N_a$
and Casimir energy $\E_a$ (conds. \mbox{(I-5.8,9)}), and then aim
at obtaining finite limits of these values ``per unit area'' when
the region increases to the whole plane. We shall not investigate
the problem of the thermodynamic limit in full generality and
restrict attention to the following cases. The modified space
$\K_\bot$ is a Hilbert space of functions on a rectangle with
sides $(L_x,L_y)$, the modified operator $h_\bot$ has an
orthonormal basis of eigenvectors:
\begin{equation}\label{sc-eig}
 h_\bot\psi_{kl}=\ep_{kl}\psi_{kl}\,,\quad
 \ep_{kl}=\sqrt{(k\ep_x)^2+(l\ep_y)^2}\,,
 \quad
 \ep_x=\pi/L_x\,,\quad \ep_y=\pi/L_y\,,
\end{equation}
and the scope of the values of $(k,l)$ is either
\begin{equation}\label{sc-cases}
 \text{(D)}\quad \mN\times\mN\,,\qquad \text{or}\quad
 \text{(N)}\quad
 (\mN\times\mN)\cup(\{0\}\times\mN)\cup(\mN\times\{0\})\,.
\end{equation}
For the scalar field the following choices lead to these two
cases:
\begin{itemize}
 \item[(D)] $\K_\bot=L^2(\<-L_x/2,L_x/2\rangle\times\<-L_y/2,L_y/2\rangle,dx\,dy)$ and
$-h^2_\bot$ is the two-dimen\-sional Laplace operator with the
Dirichlet boundary conditions;
 \item [(N)] $\K_\bot$ is the orthogonal
complement of constant functions in the space\\
$L^2(\<-L_x/2,L_x/2\rangle\times\<-L_y/2,L_y/2\rangle,dx\,dy)$,
and $-h^2_\bot$ is the two-dimensional Laplace operator with the
Neumann boundary conditions.
\end{itemize}
In the case of Neumann boundary conditions we exclude constants,
as they would lead to more singular case. Related, but slightly
different choices, also lead to the cases (D) and (N) for the
electromagnetic field (see Section \ref{el}). For this field the
absence of constants appears naturally.

With these modifications we can now introduce external conditions.
As a~result the $z$-motion Hamiltonian $h_z$ is replaced by a new
positive operator $h_{za}$, and in analogy to Eq.\,(\ref{sc-h}) we
have
\begin{equation}\label{sc-ha}
 \D(h_a)=\D(h_\bot\otimes\id)\cap\D(\id\otimes h_{za})\,,\quad
 h_a=\sqrt{(h_\bot\otimes\id)^2+(\id\otimes h_{za})^2}\,.
\end{equation}
A particular choice of $h_{za}$ will be proposed later, but first
we want to identify all models fitting into the scheme, i.e. all
perturbations which are admissible in the sense described in
\cite{AH}. A preliminary condition for this is that the symplectic
mapping $L_a$ determined by $h$ and $h_a$ as in Section I-5 be
bounded, with bounded inverse -- condition (i) in that section. In
the rest of the present section we find an equivalent form of this
condition in terms of the $z$-motion Hamiltonians $h_z$ and
$h_{za}$, and express quantities $\N_a$ and $\E_a$ in terms of
these operators. In the next two sections the conditions for
finiteness of $\N_a$ and $\E_a$ and their infinite plane limits
per area are obtained.

First of all we observe that each of the subspaces
$\psi_{kl}\otimes\K_z$ is invariant under both operators $h$ and
$h_a$. If we fix the basis $\{\psi_{kl}\}$ (we give up the freedom
of phase factor multiplication) then each of the subspaces
$\psi_{kl}\otimes\K_z$ is naturally unitarily isomorphic to
$\K_z$. It is then easy to see that if we denote for $u\geq0$
\begin{align}
 &\D(h(u))=\D(h_z)\,, & & h(u)=\sqrt{h^2_z+u\id}\,,\label{sc-hu}\\
 &\D(h_a(u))=\D(h_{za})\,,& &
 h_a(u)=\sqrt{h^2_{za}+u\id}\,,\label{sc-hua}
\end{align}
then also
\begin{equation}
 \D(h^{1/2}(u))=\D(h_z^{1/2})\,, \quad
 \D(h_a^{1/2}(u))=\D(h_{za}^{1/2})
\end{equation}
and one can use the following identifications
\begin{equation}
 \K=\bigoplus_{kl}(\K_z)_{kl}\,,\quad
 h^{1/2}=\bigoplus_{kl} h^{1/2}(\ep_{kl}^2)\,,\quad
 h_a^{1/2}=\bigoplus_{kl} h_a^{1/2}(\ep_{kl}^2)\,,
\end{equation}
where $(\K_z)_{kl}$ are identical copies of $\K_z$. The spectrum
of both $h^{1/2}(u)$ and $h_a^{1/2}(u)$ is contained in
$\<u^{1/4},\infty)$, so both operators $h^{-1/2}$ and $h_a^{-1/2}$
are bounded, with $\|h_z^{-1/2}\|,\|h_a^{-1/2}\|\leq
(\min_{kl}\{\ep_{kl}\})^{-1/2}$.

After these preliminaries it is now easy to show that the
following conditions are equivalent:
\begin{itemize}
 \item[(i)] The operators $h$ and $h_a$ (cf.\ (\ref{sc-h}) and
(\ref{sc-ha}) resp.) satisfy the conditions
\begin{gather}
 \D(h_a^{\pm1/2})=\D(h^{\pm1/2})\,,\label{sc-i1}\\
 B_a\equiv h_a^{1/2}h^{-1/2}\ \text{and}\ B_a^{-1}\ \text{extend
 to bounded operators in}\ \K\,.\label{sc-i2}
\end{gather}
\item[$\mr{(i)'}$] The operators $h_z$ and $h_{za}$ satisfy the
conditions
\begin{gather}
 \D(h_{za}^{1/2})=\D(h_z^{1/2})\,,\label{sc-ii1}\\
 \begin{split}
 &B_a(u)\equiv h_a^{1/2}(u)h^{-1/2}(u)\ \text{and}\ B_a^{-1}(u)\
 \text{are uniformly}\\
  &\text{bounded on each set}\ u\in\<v,\infty)\,,\ v>0\,.
  \end{split}\label{sc-ii2}
\end{gather}
\end{itemize}

Note that the condition (i) is identical with the condition (ii)
of Section I-5, which is equivalent to the symplectic mapping
$L_a$ being bounded together with its inverse (condition (i) in
Sec.\ I-5).

To prove the equivalence suppose first that $\mr{(i)'}$ holds.
Then
\begin{equation}
 B_a=\bigoplus_{kl}B_a(\ep^2_{kl})\,,
\end{equation}
so
$\dsp\|B^{\pm1}_a\|\leq\max_{kl}\|B^{\pm1}_a(\ep^2_{kl})\|\leq\con$.
Then $h_a^{1/2}=B_ah^{1/2}$ and $h^{1/2}=B_a^{-1}h^{1/2}_a$, so
$h^{1/2}$ and $h_a^{1/2}$ have a common domain. As $h^{-1/2}$ and
$h_a^{-1/2}$ are bounded, this ends the proof of (i). Conversely,
suppose that (i) is true. Then the restrictions of the domains of
$h^{1/2}$ and $h_a^{1/2}$ to $(\K_z)_{kl}$ must be equal, which
implies Eq.\,(\ref{sc-ii1}). Condition (\ref{sc-i2}) implies that
$B^{\pm1}_a(\ep^2_{kl})$ are uniformly bounded. But for $u,w>0$
\begin{equation}
 B_a(u)=h_a^{1/2}(u)h_a^{-1/2}(w)B_a(w)h^{1/2}(w)h^{-1/2}(u)\,,
\end{equation}
so by spectral analysis for $h$ and $h_a$ one finds
\begin{equation}
 \|B^{\pm1}_a(u)\|\leq
 \max\bigg\{\Big(\frac{u}{w}\Big)^{1/4},\Big(\frac{w}{u}\Big)^{1/4}\bigg\}\,
 \|B^{\pm1}_a(w)\|\,.
\end{equation}
Setting here $w=\ep^2_{kl}$ and $u$ in the interval between
$\ep^2_{kl}$ and a neighbouring value of $\ep^2_{..}$, or between
zero and $\ep^2_{kl}$, if the latter is the minimal value, one
shows that the condition (\ref{sc-ii2}) is satisfied.

The equivalence shows that if (i) is satisfied for some $L_x,L_y$,
then it is true for all finite values of these parameters.

If the equivalent conditions (i) and $\mr{(i)'}$ are satisfied
then the operators $T_a$ and $S_a$ defining the Bogoliubov
transformation are bounded, and given by
\begin{equation}
 T_a=\bigoplus_{kl}T_a(\ep^2_{kl})\,,\qquad
 S_a=\bigoplus_{kl}S_a(\ep^2_{kl})\,,
\end{equation}
where
\begin{equation}\label{sc-sutu}
 T_a(u)=\frac{1}{2}\big[B_a^{-1}(u)+B_a^*(u)\big]\,,\qquad
 S_a(u)=\frac{1}{2}\big[B_a^{-1}(u)-B_a^*(u)\big]K\,,
\end{equation}
and $K$ is the operator of complex conjugation. The quantities
$\N_a$ (particle number) and $\E_a$ (Casimir energy), which have
to be investigated, may be now written as
\begin{align}
 &\begin{aligned}
 &\N_a\equiv \Tr[S_aS^*_a]=\sum_{kl}\Tr\big[N_a(\ep^2_{kl})\big]\,,\\
 &\hspace{4em}
 \text{where}\hspace{3em} N_a(u)=S_a(u)S_a^*(u)\,,
 \end{aligned}\label{sc-N}\\
 &\begin{aligned}
 &\E_a\equiv \Tr\big[h^{1/2}S_aS^*_ah^{1/2}\big]=\sum_{kl}\Tr
  \big[E_a(\ep^2_{kl})\big]\,,\\
 &\hspace{4em}\text{where}\hspace{3em}E_a(u)
 =h^{1/2}(u)S_a(u)S_a^*(u)h^{1/2}(u)\,,
 \end{aligned}\label{sc-E}
\end{align}
where we do not know yet whether these expressions are finite (but
they are unambiguously defined as positive numbers, finite or not,
as the operators are positive).

\setcounter{equation}{0}

\section{The necessary and sufficient condition for the unitary
equivalence of representations and the existence of finite limit
$\dsp\boldsymbol{n_a\equiv\lim_{L_x,L_y\to\infty}\N_a/L_xL_y}$}\label{eq}

We now add the condition:
\begin{itemize}
 \item[(ii)] The ground state representations
determined by $h$ and $h_a$ are unitarily equivalent, and the
infinite plane limit of particle number per area is well
defined,~i.e.
\begin{equation}\label{eq-cond}
 \N_a<\infty\quad \text{for all}\ L_x,L_y\,,\qquad
 \exists\ \lim_{L_x,L_y\to\infty}\frac{\N_a}{L_xL_y}<\infty\,.
\end{equation}
\end{itemize}

We show in this section that the conditions (i) and (ii) are
satisfied if, and only if:
\begin{itemize}
 \item[(Eq)] The operators $h_z$ and $h_{za}$ have a common domain and
$h_{za}-h_z$ extends to a Hilbert-Schmidt operator on $\K_z$, i.e.
\begin{equation}\label{eq-ctr}
 \Tr(h_{za}-h_z)^2<\infty\,.
\end{equation}
\end{itemize}
 In this case
\begin{equation}\label{eq-eq}
 n_a\equiv\lim_{L_x,L_y\to\infty}\frac{\N_a}{L_xL_y}
 =\frac{1}{16\pi}\Tr(h_{za}-h_z)^2\,.
\end{equation}
In the case of Dirichlet boundary conditions for $h_\bot$ (case
(D) in Eq.\,(\ref{sc-cases})) the limit in Eqs.\,(\ref{eq-cond})
and (\ref{eq-eq}) can be taken in arbitrary way, but in the case
of Neumann conditions (case (N) in Eq.\,(\ref{sc-cases})) one has
to keep $1/M\leq L_x/L_y\leq M$ for some arbitrary, but fixed
$M>1$ (which means that the limit takes place over the values of
$(L_x,L_y)$ within a conic proper subset of
$(0,\infty)\times(0,\infty)$).

\subsection{(Eq) $\boldsymbol{\then}$ (i) and (ii)}

Suppose first that the condition (Eq) is true, and denote
$c=\|h_{za}-h_z\|$. Then for $\vp\in\D(h_{za})=\D(h_z)$ we have
\begin{multline*}
 \|h_a(u)\vp\|^2=\|h_{za}\vp\|^2+u\|\vp\|^2
 \leq(\|h_z\vp\|+c\|\vp\|)^2+u\|\vp\|^2\\
 <\bigg(1+\frac{c}{\sqrt{u}}+\frac{c^2}{u}\bigg)
 (\|h_z\vp\|^2+u\|\vp\|^2)
 =\bigg(1+\frac{c}{\sqrt{u}}+\frac{c^2}{u}\bigg)
 \|h(u)\vp\|^2\,.
\end{multline*}
Now we use the monotonicity of the square root: if $A$ and $B$ are
positive operators, $\D(A)\subseteq\D(B)$ and $\|B\psi\|\leq
\|A\psi\|$ for all $\psi\in\D(A)$, then also
$\D(A^{1/2})\subseteq\D(B^{1/2})$ and
$\|B^{1/2}\psi\|\leq\|A^{1/2}\psi\|$ for all $\psi\in\D(A^{1/2})$
(see \cite{rs}, proof of Thm.\,X.18). Thus we get
\begin{equation*}
 \|h_a^{1/2}(u)\vp\|<
 \bigg(1+\frac{c}{\sqrt{u}}+\frac{c^2}{u}\bigg)^{1/4}
 \|h^{1/2}(u)\vp\|\,
\end{equation*}
This implies $\|B_a(u)\|<\big(1+c/\sqrt{u}+c^2/u\big)^{1/4}$.
Changing the roles of $h(u)$ and $h_a(u)$ one exhausts the
condition $\mr{(i)'}$.

We introduce a bounded selfadjoint operator on $\K_z$:
\begin{equation}
 \Delta_a(u)=h_a(u)-h(u)
\end{equation}
(boundedness by $\|h_a(u)-h_a(0)\|\leq\sqrt{u}$,
$\|h(u)-h(0)\|\leq\sqrt{u}$ and the boundedness of $h_{za}-h_z$).
In the further course of the proof we shall need several theorems
on Hilbert-Schmidt properties of operators like $\Delta_a(u)$ and
$S_a(u)$. As their demonstration is somewhat more technical we
shift their derivation to Appendix \ref{hs} and in the main text
use the results. We also remind the reader the following facts on
operators in Hilbert space: if $B$ is bounded and $H$ is
Hilbert-Schmidt, then $BH$ and $HB$ are Hilbert-Schmidt, and if in
addition $K$ is also Hilbert-Schmidt, then
\[
 \Tr(BHK)=\Tr(HKB)=\Tr(KBH)\,.
\]
Furthermore, if $B$ is in addition positive, then
\[
 \Tr\big(\sqrt{B}HH^*\sqrt{B}\big)=\Tr(H^*BH)\leq\|B\|\Tr(H^*H)\,.
\]
We use these properties extensively in what follows.

We have to show that the expression (\ref{sc-N}) satisfies
condition (ii), and Eq.\,(\ref{eq-eq}) holds. We have assumed that
$\Delta_a(0)$ is a HS operator. It follows then from the result
$\mr{(iv)_B}$ of Appendix \ref{hs} that for all $u\geq0$ the
operators $\Delta_a(u)$ are also Hilbert-Schmidt, and the function
$\<0,\infty)\ni u\mapsto\Tr\Delta_a^2(u)$ is continuous on its
domain and continuously differentiable on $(0,\infty)$,
decreasing, and $\dsp\lim_{u\to\infty}\Tr\Delta_a^2(u)=0$.
Moreover, the differentiation on $u$ may be carried out by
formally pulling the derivative under the trace sign and
differentiating formally the $u$-dependent operators. Using the
formal rule
\begin{equation}\label{eq-derd}
  2\frac{d}{du}\Delta_a(u)=-h_a^{-1}(u)\Delta_a(u)h^{-1}(u)=
  -h^{-1}(u)\Delta_a(u)h_a^{-1}(u)
\end{equation}
and the identity
\begin{equation}
 S_a(u)=-\frac{1}{2}h^{-1/2}(u)\Delta_a(u)h_a^{-1/2}(u)K
\end{equation}
 following from the formula (\ref{sc-sutu}) one finds
\begin{equation}\label{eq-dn}
 \begin{split}
 2\frac{d}{du}\Tr\Delta_a^2(u)=
 &-\Tr\big[h^{-1}(u)\Delta_a(u)h_a^{-1}(u)\Delta_a(u)\big]\\
 &-\Tr\big[\Delta_a(u)h_a^{-1}(u)\Delta_a(u)h^{-1}(u)\big]
 =-8\Tr N_a(u)\,.
  \end{split}
\end{equation}
It is now clear that the function
 $(0,\infty)\ni u\mapsto\Tr N_a(u)$
is continuous, positive, and
\begin{equation}\label{eq-intn}
 \int_0^\infty\Tr N_a(u)\,du
 =\frac{1}{4}\Tr\Delta_a^2(0)<\infty\,.
\end{equation}
Again by the result $\mr{(iv)_B}$ of Appendix \ref{hs} the
function $(0,\infty)\ni u\mapsto\Tr N_a(u)$ is decreasing. With
the use of formulas (\ref{sum-ineq'}) and (\ref{sum-si2}) of
Appendix \ref{sum} this is sufficient to conclude that in the case
(D) of Eq.\,(\ref{sc-cases}) the condition (\ref{eq-cond}), and
Eq.\,(\ref{eq-eq}) are satisfied. In the case (N) of
Eq.\,(\ref{sc-cases}) $\N_a$ is bigger by
\[
 \N_a^{(N)}-\N_a^{(D)}=\dsp\ep_x\ep_y\sum_{l=1}^\infty\Tr
 N_a((l\ep_y)^2)+\{x\leftrightarrow y\}\,.
\]
 If one keeps $1/M\leq
\ep_x/\ep_y\leq M$ then
\begin{equation*}
 \N_a^{(N)}-\N_a^{(D)}\leq
 M\ep^2_y\sum_{l=1}^\infty\Tr N_a((l\ep_y)^2)
 +M\ep^2_x\sum_{l=1}^\infty\Tr N_a((l\ep_x)^2)\,,
\end{equation*}
which by (\ref{eq-intn}) and the properties (\ref{sum-li3}) and
(\ref{sum-ineq2}) of Appendix \ref{sum} is finite and tends to
zero for $\ep_x,\ep_y\to 0$. Thus in both cases (D) and (N) the
condition (ii) and the equation (\ref{eq-eq}) follow from the
criterion (\ref{eq-ctr}).

\subsection{(i) and (ii) $\boldsymbol{\then}$ (Eq)}

We now turn to the proof that the criterion (Eq) follows from (i)
and (ii). This fact is of interest, as it shows that in the
context adopted in this paper there is no escape from this
condition. Suppose that the first of conditions (\ref{eq-cond}) is
satisfied. By the result $\mr{(iv)_B}$ of Appendix \ref{hs} if
$\Tr N_a(u)$ is finite for any $u>0$, then it is finite for all
$u\in(0,+\infty)$, and the function $(0,+\infty)\ni u\mapsto \Tr
N_a(u)$ is continuously differentiable and decreasing. Thus if
$\N_a$ is finite for all $L_x,L_y$, then by the first of
inequalities in (\ref{sum-ineq'}) we have for each $v>0$:
\begin{equation}\label{eq-ib}
  \int_v^\infty\Tr N_a(u)\,du<\infty\,.
\end{equation}

Let now $P_\beta=P_{\<0,\beta\rangle}(h_z)$ be the projection
operator from the spectral family of the operator $h_z$,
projecting onto the spectral interval $\<0,\beta\rangle$, and
similarly \mbox{$P_{a\gamma}=P_{\<0,\gamma\rangle}(h_{za})$} for
the operator $h_{za}$. Then for each $u>0$ the expression
$P_\beta\Delta_a(u)P_{a\gamma}$ defines a bounded operator
(although we do not know yet whether $\Delta_a(u)$ alone makes
sense), and
\begin{multline}\label{}
  \Tr[P_\beta\Delta_a(u)P_{a\gamma}\Delta_a(u)P_\beta]\\
  \leq
  4\sqrt{(\beta^2+u)(\gamma^2+u)}
  \Tr[P_\beta S_a(u)P_{a\gamma}S^*_a(u)P_\beta]\\
  \leq 4\sqrt{(\beta^2+u)(\gamma^2+u)}\Tr N_a(u)\to0\quad
  (u\to\infty)\,,
\end{multline}
the last line by the property (\ref{sum-li2}) (note that here and
in what follows all operators under the trace sign are positive).
It follows from the result $\mr{(iv)_B}$ of Appendix \ref{hs} that
the function on the l.h.\ side may be differentiated as that in
Eq.\,(\ref{eq-dn}), so one finds
\begin{equation}\label{eq-derpd}
  \frac{d}{du}\Tr[P_\beta\Delta_a(u)P_{a\gamma}\Delta_a(u)P_\beta]
  =-4\Tr[P_\beta S_a(u)P_{a\gamma}S^*_a(u)P_\beta]\,.
\end{equation}
Integrating this identity on $\<v,\infty)$, $v>0$,  one has
\begin{equation}\label{eq-idp}
  \Tr[P_\beta\Delta_a(v)P_{a\gamma}\Delta_a(v)P_\beta]
  =4\int_{v}^\infty
  \Tr[P_\beta S_a(u)P_{a\gamma}S^*_a(u)P_\beta]\,du\,.
\end{equation}
The integrand on the r.h.\ side is bounded by $\Tr N_a(u)$, so the
double limit \mbox{$\beta,\gamma\to\infty$} exists. Thus
\begin{equation}\label{eq-trd}
  \Tr\Delta_a^2(v)=4\int_v^\infty\Tr N_a(u)\,du\,.
\end{equation}
Now we let the second of the assumptions in (\ref{eq-cond}) come
into play. By formula (\ref{sum-si2}) this implies that the r.h.\
side of the last formula has a finite limit for $v\to 0$. As a
result we have
\begin{equation}\label{eq-lim}
  \Tr\Delta_a^2(v)\leq 4\int_0^\infty\Tr N_a(u)\,du<\infty\,.
\end{equation}
By the result $\mr{(iv)_B}$ this implies that $\Delta_a(0)$ is HS,
which completes the proof that (\ref{eq-ctr}) is a necessary
condition.

\setcounter{equation}{0}

\section{The necessary and sufficient condition for finite
backreaction energy and its limit\\
$\dsp\boldsymbol{\vep_a\equiv\lim_{L_x,L_y\to\infty}\E_a/L_xL_y}$}\label{en}

We now add the energy condition:
\begin{itemize}
\item[(iii)] The backreaction energy $\E_a$ is finite for all
$L_x,L_y$, and the limit of Casimir energy per area is well
defined for infinite plane limit:
\begin{equation}\label{en-conden}
  \E_a<\infty\,,\qquad
  \exists\ \lim_{L_x,L_y\to\infty}\frac{\E_a}{L_xL_y}<\infty\,.
\end{equation}
\end{itemize}

We show in this section that if the conditions (i) and (ii) (or
criterion (Eq)) are satisfied, then (iii) is fulfilled if, and
only if:
\begin{itemize}
\item[(En)] The operator $(h_{za}-h_z)h_z^{1/2}$ is a
Hilbert-Schmidt operator, that is
\begin{equation}\label{en-ctren}
  \Tr[(h_{za}-h_z)h_z(h_{za}-h_z)]<\infty\,.
\end{equation}
\end{itemize}
 If this condition is satisfied, then
\begin{equation}\label{en-energy}
 \vep_a\equiv\lim_{L_x,L_y\to\infty}\frac{\E_a}{L_xL_y}
 =\frac{1}{24\pi}\Tr[(h_{za}-h_z)(2h_z+h_{za})(h_{za}-h_z)]\,.
\end{equation}
The limit with respect to $L_x,L_y$ is specified as in
(\ref{eq-eq}).

\subsection{(Eq) and (En) $\boldsymbol{\then}$ (iii)}

The proof is closely analogous to the one in the last section.
Suppose that criterions (Eq) and (En) are satisfied. We have to
show that the expression (\ref{sc-E}) satisfies the conditions
(\ref{en-conden}) and Eq.\,(\ref{en-energy}). We have assumed that
$\Tr[\Delta_a(0)^2]<\infty$ and
\mbox{$\Tr[\Delta_a(0)h(0)\Delta_a(0)]<\infty$}. It follows then
from the result $\mr{(v)_B}$ of Appendix~\ref{hs} that the
functions $u\mapsto\Tr[\Delta_a(u)h(u)\Delta_a(u)]$ and
$u\mapsto\Tr[\Delta_a(u)h_a(u)\Delta_a(u)]$ are continuous on
$\<0,\infty)$ and continuously differentiable on $(0,\infty)$, and
tend to zero for $u\to\infty$. Also, the formal differentiation
with respect to $u$ as in Eqs.\,(\ref{eq-derd}), (\ref{eq-dn}),
yields the correct result. A direct calculation then yields
\begin{equation}\label{en-dener}
  \frac{1}{6}\,\frac{d}{du}\Big\{2\Tr[\Delta_a(u)h(u)\Delta_a(u)]
  +\Tr[\Delta_a(u)h_a(u)\Delta_a(u)]\Big\}=-\Tr E_a(u)\,,
\end{equation}
and the integration leads to
\begin{equation}\label{en-inten}
  \int_0^\infty\Tr E_a(u)\,du=\frac{1}{6}
  \Tr[\Delta_a(0)(2h(0)+h_a(0))\Delta_a(0)]<\infty\,.
\end{equation}
By the result $\mr{(iv)_B}$ the function $\Tr E_a(u)$ is
non-increasing. By a reasoning analogous to that following
Eq.\,(\ref{eq-intn}) one shows that $\E_a^{(N)}-\E_a^{(D)}$
vanishes in the limit. One concludes that the conditions
(\ref{en-conden}) and the equation (\ref{en-energy}) are
satisfied.

\subsection{(Eq) and (iii) $\boldsymbol{\then}$ (En)}

Conversely, suppose that $h_{za}-h_z$ is a Hilbert-Schmidt
operator and $\E_a$ is finite. Similarly as in the case of $\Tr
N_a(u)$ it follows from the result $\mr{(iv)_B}$ of Appendix
\ref{hs} that if $\Tr E_a(u)$ is finite for any $u>0$, then it is
finite for all $u\in(0,+\infty)$, and the function $(0,+\infty)\ni
u\mapsto \Tr E_a(u)$ is continuously differentiable and
decreasing. Thus if $\E_a$ is finite, then for each $v>0$
\begin{equation}\label{sc-iben}
  \int_v^\infty\Tr E_a(u)\,du<\infty\,.
\end{equation}
Therefore
\begin{equation}\label{sc-pen}
 \begin{split}
  \Tr[&P_\beta\Delta_a(u)P_{a\gamma}
  h_a(u)P_{a\gamma}\Delta_a(u)P_\beta]\\
  &\leq  (\gamma^2+u)\Tr\Big[P_\beta\Delta_a(u)
  \frac{P_{a\gamma}}{h_a(u)}\Delta_a(u)P_\beta\Big]
  \leq(\gamma^2+u)\Tr E_a(u)<\infty\,,
 \end{split}
\end{equation}
\begin{equation}
 \begin{split}
  \Tr[&P_{a\gamma}\Delta_a(u)P_\beta
  h(u)P_\beta\Delta_a(u)P_{a\gamma}]\\
  &\leq  \sqrt{(\beta^2+u)(\gamma^2+u)}
  \Tr\Big[\frac{P_{a\gamma}}{h_{za}^{1/2}(u)}\Delta_a(u)
  P_\beta\Delta_a(u)\frac{P_{a\gamma}}{h_a^{1/2}(u)}\Big]\\
  &\hspace{5cm}\leq\sqrt{(\beta^2+u)(\gamma^2+u)}
  \Tr E_a(u)<\infty\,,
 \end{split}
\end{equation}
and by (\ref{sum-li2}) both functions tend to zero for
$u\to\infty$. Differentiation is again allowed and gives
\begin{multline}\label{sc-diffen}
  \frac{1}{6}\,\frac{d}{du}\Big\{2
  \Tr[P_{a\gamma}\Delta_a(u)P_\beta
  h(u)P_\beta\Delta_a(u)P_{a\gamma}]+
  \Tr[P_\beta\Delta_a(u)P_{a\gamma}
  h_a(u)P_{a\gamma}\Delta_a(u)P_\beta]\Big\}\\
  =-\Tr\Big[P_\beta\Delta_a(u)
  \frac{P_{a\gamma}}{h_a(u)}\Delta_a(u)P_\beta\Big]\,.
\end{multline}
Now we take into the consideration the second condition in
(\ref{en-conden}) and go over the steps analogous to those
following Eq.\,(\ref{eq-derpd}). This gives
\begin{equation}\label{sc-enlin}
  \Tr[\Delta_a(v)h(v)\Delta_a(v)]\leq
  3\int_0^\infty\Tr E_a(u)\,du<\infty\,.
\end{equation}
This implies, by $\mr{(v)_B}$, that $\Delta_a(0)h^{1/2}(0)$ is
also HS, which ends the proof.

\setcounter{equation}{0}

\section{Modified Dirichlet and Neumann conditions}\label{mdn}

From now on we can restrict attention to the $z$-motion dynamics,
and specify a class of models. We set $\K_z=L^2(\mR)$,
$h_z^2=-\p_z^2$, where the unique selfadjoint extension of the
second derivative defined for $\mathcal{S}(\mR)$ is meant. We want
the external conditions to be some modification of the strict
boundary conditions enforced on the planes $z=\pm b$, where
$2b=a$. The Hamiltonian of the $z$-motion for such conditions is
determined by $(\hb)^2=-(\p^2_z)_B$, where $B$ stands for
Dirichlet or Neumann conditions at $z=\pm b$. Let $F$ and $G$ be
real, non-negative, bounded, measurable functions on
$\<0,\infty)$. We postulate a class of models by setting
\begin{equation}\label{mdn-mod}
    h_{za}=h_z+G(h_z)\big[F(\hb)-F(h_z)\big]G(h_z)\,.
\end{equation}
We assume that for all $u\geq0$ we have $F(u)\leq u$ and $G(u)\leq
1$, which guarantees the positivity of $h_{za}$. The strict
boundary conditions are recovered by formally setting $F(u)=u$ and
$G(u)=1$ for all $u\geq0$. Our intention is to keep these
equalities for $u$ ``not to large'', and modify $F$ and $G$ so
that for $u\to\infty$ $F$ remains bounded, and $G$ (possibly)
tends to zero. This seems to model correctly the idea that the
boundaries should become transparent for very energetic particles.
We show in this section the following:

\begin{itemize}
 \item[(Mod)] Let the functions $F\in\C^2\big(\<0,\infty)\big)$ and
 $G\in\C\big(\<0,\infty)\big)$ satisfy the estimates
\begin{equation}
 \begin{aligned}
 &0\leq F(p)\leq\con\,,\quad F(p)\leq p\,,\quad
 0\leq G(p)\leq 1\,,\\
 &G(p)\leq\con (p+1)^{-\alpha}\,,\quad
 |F^{(2)}(p)|\leq\con\,(p+1)^{-2-\gamma}\\
 &\text{for some}\
 \alpha,\gamma\in(0,1)\ \text{and such that}\
 4\alpha+2\gamma>1\,.
\end{aligned}\label{mdn-fgb}
\end{equation}
Then $h_{za}-h_z$, where $h_{za}$ is the modified $z$-motion
operator given by (\ref{mdn-mod}), satisfies conditions (Eq) and
(En). In this case $h_{za}-h_z$ is
 an integral operator
\begin{equation}
    (\psi,[h_{za}-h_z]\psi')=\int
    \ov{\wh{\psi}(p)}\,K^{F,G}_a(p,p')\,
    \wh{\psi'}(p')\,dpdp'\,,\label{mdn-GFF}
\end{equation}
with $K^{F,G}_a(p,p')\in L^2(\mR^2)$ and
$\sqrt{|p|}K^{F,G}_a(p,p')\in L^2(\mR^2)$. The explicit form of
$K^{F,G}_a$ is obtained in (\ref{mdn-kerF} -- \ref{mdn-kerFG})
below. The Casimir energy is given by
\begin{equation}\label{mdn-ce}
 \begin{split}
    \vep_a&=\frac{1}{24\pi}\Tr
    \Big\{(h_{za}-h_z)[3h_z+(h_{za}-h_z)](h_{za}-h_z)\Big\}\\
    &=\frac{1}{24\pi}
    \bigg\{\int 3|p|[K^{F,G}_a(p,p')]^2\,dp\,dp'\\
    &\hspace{40pt}+\int K^{F,G}_a(p,p')K^{F,G}_a(p',p'')K^{F,G}_a(p'',p)
    \,dp\,dp'\,dp''\bigg\}\,.
 \end{split}
\end{equation}
Moreover, let for $\mu>0$:
\begin{equation}\label{mdn-scaled}
    F_\mu(p)=\mu F(p/\mu)\,,\quad
    G_\mu(p)=G(p/\mu)\,,
\end{equation}
and denote by $\vep^\mu_a$ the Casimir energy for the model with
$F,G$ replaced with $F_\mu,G_\mu$. Then
\begin{equation}\label{mdn-scaling}
    \vep^\mu_a=\mu^3\vep_{\mu a}\,.
\end{equation}
\end{itemize}

\noindent As explained in the opening paragraph of this section,
we believe that (\ref{mdn-mod}), with $F$ and $G$ in suitable
classes, defines a class of models reasonably imitating boundary
conditions for low energies, while giving the means for
``softening'' of the walls. More specific of the assumptions on
$F$ and $G$ are technical. (Mod) then shows that indeed those
models are not only admissible for the discussion of adiabatic
backreaction as explained in earlier sections, but also
susceptible to a rigorous treatment. Low energy behaviour
$F(p)\sim p$ and $G(p)\sim 1$ for ``small''~$p$, important for the
above interpretation, is not needed in (Mod). Once these
assumptions are added in the next section, the scaling properties
described in (\ref{mdn-scaled}) and (\ref{mdn-scaling}) provide
means for investigation of the sharp boundary limit.

\subsection{Spectral representation of $\boldsymbol{\hb}$}

The spectral representation for $h_z$ is obtained by Fourier
transforming $\psi\mapsto\hp$ ($\psi,\hp\in L^2(\mR)$), so that in
the new representation $h_z$ acts by $\wh{h_z\psi}(p)=|p|\hp(p)$,
where the conventions for the transform are defined by
\begin{equation}\label{mod-ft}
    \hp(p)=\frac{1}{\sqrt{2\pi}}\int\psi(z)e^{-ipz}dz\,.
\end{equation}
 To prove
(Mod) it is sufficient to show that in this representation
\mbox{$F(h_{za}^B)-F(h_z)$} is an integral operator with a kernel
$K_a^F(p,p')$ such that both functions $G(|p|)K_a^F(p,p')G(|p'|)$
and $\sqrt{|p|}G(|p|)K_a^F(p,p')G(|p'|)$ are in $L^2(\mR^2)$.

We start by identifying the spectral representation for $\hb$. It
will be convenient to denote
\begin{equation}\label{mdn-d}
    \delta=\begin{cases}-1&\text{Dirichlet case}\,,\\
    +1&\text{Neumann case}\,,\end{cases}
\end{equation}
and introduce
\begin{equation}\label{mdn-ld}
  L^2_\delta=\{\chi\in L^2(\mR)\mid
  \chi(-z)=\delta\chi(z)\}\,,\quad
  l^2_\delta=\{\chi\in l^2(\mZ)\mid \chi(-k)=\delta\chi(k)\}\,.
\end{equation}
For $\psi\in L^2(\mR)$ (in the initial position representation) we
denote
\begin{equation}
  \psi_-(z)=\theta(-z-b)\psi(z)\,,\quad
  \psi_0(z)=\theta(z+b)\theta(-z+b)\psi(z)\,,\quad
  \psi_+(z)=\theta(z-b)\psi(z)\,.
\end{equation}
Then the following rule defines a unitary transformation:
\begin{equation}\label{mdn-uni}
 \begin{split}
    L^2(\mR)\ni\psi&\mapsto\Psi\equiv(\wh{\psi^\da_-},\wh{\psi^\da_0},\wh{\psi^\da_+})\in
    L^2_\da\oplus l^2_\da\oplus L^2_\da\,,\\
    \wh{\psi^\da_-}(r)&=\frac{1}{\sqrt{2}}
    \lp e^{-ibr}\wh{\psi_-}(r)+\da\, e^{+ibr}\wh{\psi_-}(-r)\rp\,,\\
    \wh{\psi^\da_0}(k)&=\sqrt{\frac{\pi}{2a}}
    \lp i^k\wh{\psi_0}(k\ep)+\da\, i^{-k}\wh{\psi_0}(-k\ep)\rp\,,\quad
    \ep=\frac{\pi}{a}\,,\\
    \wh{\psi^\da_+}(r)&=\frac{1}{\sqrt{2}}
    \lp e^{+ibr}\wh{\psi_+}(r)+\da\, e^{-ibr}\wh{\psi_+}(-r)\rp\,,
 \end{split}
\end{equation}
where $(\wh{\psi_-},\wh{\psi_0},\wh{\psi_+})$ are Fourier
transforms of $(\psi_-,\psi_0,\psi_+)$ respectively. In this new
representation operator $\hb$ acts on $\wh{\psi^\da_\pm}(r)$ as
multiplication by $|r|$, and on $\wh{\psi^\da_0}(k)$ as
multiplication by $|k|\ep$. This is easily seen by using an
overcomplete set of eigenfunctions in the three regions
\begin{equation}\label{mdn-eig}
 \begin{split}
    \vp_\pm(r,z)&=\frac{1}{2\sqrt{\pi}}
    (e^{ir(z\mp b)}+\da e^{-ir(z\mp b)})=\da\,\vp_\pm(-r,z)\,,\quad
    z\gtrless\pm b\\
    \vp_0(k,z)&=\frac{1}{2\sqrt{a}}
    (i^{-k}e^{ik\ep z}+\da i^k e^{-ik\ep
    z})=\da\,\vp_0(-k,z)\,,\quad
    z\in\<-b,b\rangle
 \end{split}
\end{equation}
normalized by
\begin{equation}\label{mdn-norm}
 \begin{split}
    \int\ov{\vp_\pm(r,z)}\vp_\pm(r',z)dz&=\frac{1}{2}[\da(r-r')+\da\,\da(r+r')]\,,\\
    \int\ov{\vp_0(k,z)}\vp_0(k',z)dz&=\frac{1}{2}[\da_{k,k'}+\da\,\da_{k,-k'}]\,.
 \end{split}
\end{equation}
We also note for later use that for $\psi\in\mathcal{S}(\mR)$ by
standard Fourier methods
\begin{equation}\label{mdn-pft}
 \begin{split}
   \wh{\psi_0}(r)&=\frac{1}{\pi}\int
   \frac{\sin b(p-r)}{p-r}\,\hp(p)\,dp\,,\\
   \wh{\psi_-}(r)&=\frac{-i}{2\pi}\int
   \frac{e^{ib(r-p)}\,\hp(p)}{p-r-i0}\,dp\,,\quad
   (\wh{\psi_-}+\wh{\psi_0})(r)=\frac{-i}{2\pi}\int
   \frac{e^{-ib(r-p)}\,\hp(p)}{p-r-i0}\,dp\,,\\
   \wh{\psi_+}(r)&=\frac{i}{2\pi}\int
   \frac{e^{-ib(r-p)}\,\hp(p)}{p-r+i0}\,dp\,,\quad
   (\wh{\psi_+}+\wh{\psi_0})(r)=\frac{i}{2\pi}\int
   \frac{e^{ib(r-p)}\,\hp(p)}{p-r+i0}\,dp\,.
 \end{split}
\end{equation}

\subsection{The integral kernel of $\boldsymbol{h_{za}-h_z}$}

Let $\psi,\psi'\in\mathcal{S}(\mR)$. Using the spectral
representation (\ref{mdn-uni}) one finds
\begin{equation}\label{mdn-F}
  \begin{split}
    (\psi,F(\hb)\psi')&=
    \int F(|r|)\left[\ov{\wh{\psi_-}(r)}\wh{\psi'_-}(r)+
    \ov{\wh{\psi_+}(r)}\wh{\psi'_+}(r)\right]\,dr\\
    &\hspace{-30pt}+\da\int
    F(|r|)\left[e^{-iar}\ov{\wh{\psi_-}(-r)}\wh{\psi'_-}(r)+
    e^{iar}\ov{\wh{\psi_+}(-r)}\wh{\psi'_+}(r)\right]\,dr\\
    &\hspace{-30pt}+\ep\sum_{k\in\mZ}F(|k|\ep)
    \ov{\wh{\psi_0}(k\ep)}\wh{\psi'_0}(k\ep)
    +\da\,\ep\sum_{k\in\mZ}F(|k|\ep)(-1)^k
    \ov{\wh{\psi_0}(-k\ep)}\wh{\psi'_0}(k\ep)
  \end{split}
\end{equation}
For further evaluation we shall need results of the Appendix
\ref{pw}. We also use the notation introduced at the beginning of
that appendix.

We represent the terms in brackets under the integral sign in the
first line by
\begin{equation}\label{}
 \begin{split}
   \ov{\wh{\psi_-}(r)}\wh{\psi'_-}(r)+
    \ov{\wh{\psi_+}(r)}\wh{\psi'_+}(r)=
    &\ov{\wh{\psi}(r)}\wh{\psi'}(r)-
    \ov{\wh{\psi_0}(r)}\wh{\psi'}(r)\\
    -&\ov{\wh{\psi_-}(r)}(\wh{\psi'_+}+\wh{\psi'_0})(r)
    -\ov{\wh{\psi_+}(r)}(\wh{\psi'_-}+\wh{\psi'_0})(r)
 \end{split}
\end{equation}
and use representation given in (\ref{mdn-pft}). Then for the
integration of the terms in the second line of the last identity
we can use identity (\ref{pw-chiint}). We get
\begin{equation}\label{mdn-1}
 \begin{split}
    \int F(|r|)&\left[\ov{\wh{\psi_-}(r)}\wh{\psi'_-}(r)+
    \ov{\wh{\psi_+}(r)}\wh{\psi'_+}(r)\right]\,dr=
    \int F(|r|)\ov{\wh{\psi}(r)}\wh{\psi'}(r)\,dr\\
    +\int&\bigg[\frac{1}{2\pi^2}\cos[b(p-p')]\chi_F(p,p')\\
    &\hspace{45pt}-\frac{\sin[b(p-p')]}{2\pi(p-p')}\big[F(|p|)+F(|p'|)\big]\bigg]
    \ov{\wh{\psi}(p)}\wh{\psi'}(p')\,dpdp'\,.
 \end{split}
\end{equation}
For the integral in the second line in (\ref{mdn-F}) we again use
(\ref{pw-chiint}) to obtain
\begin{equation}\label{mdn-2}
 \begin{split}
    \int F(|r|)&\left[e^{-iar}\ov{\wh{\psi_-}(-r)}\wh{\psi'_-}(r)+
    e^{iar}\ov{\wh{\psi_+}(-r)}\wh{\psi'_+}(r)\right]\,dr\\
    =\int&\bigg[-\frac{1}{2\pi^2}\cos[b(p-p')]\chi_F(p,-p')\\
    &\hspace{45pt}+\frac{\sin[b(p-p')]}{2\pi(p+p')}\big[F(|p|)-F(|p'|)\big]\bigg]
    \ov{\wh{\psi}(p)}\wh{\psi'}(p')\,dpdp'\,.
 \end{split}
\end{equation}
To evaluate the sums in the last line of Eq.\,(\ref{mdn-F}) we
represent $\wh{\psi_0}$ as in (\ref{mdn-pft}) and use identities
(\ref{pw-chisin}) and (\ref{pw-chiasin}), which results in
\begin{equation}
 \begin{split}
  &\ep\sum_{k\in\mZ}F(|k|\ep)
  \ov{\wh{\psi_0}(k\ep)}\wh{\psi'_0}(k\ep)
  =\int\bigg[\frac{1}{2\pi^2}\cos[b(p-p')]\chi_{F,\ep}(p,p')\\
 &\hspace{45pt}+\frac{1}{2\pi^2}\cos[b(p+p')]
 [\chi_{F,\ep}(p,p')-\chi_{F,2\ep}(p,p')]\\
 &\hspace{90pt}+\frac{\sin[b(p-p')]}{2\pi(p-p')}\big[F(|p|)+F(|p'|)\big]\bigg]
 \ov{\wh{\psi}(p)}\wh{\psi'}(p')\,dpdp'\,,\\
 \end{split}\label{mdn-3}
\end{equation}
\begin{equation}
 \begin{split}
  &\ep\sum_{k\in\mZ}F(|k|\ep)(-1)^k
  \ov{\wh{\psi_0}(-k\ep)}\wh{\psi'_0}(k\ep)
  =\int\bigg[-\frac{1}{2\pi^2}\cos[b(p-p')]\chi_{F,\ep}(p,-p')\\
  &\hspace{45pt}-\frac{1}{2\pi^2}\cos[b(p+p')]
  [\chi_{F,\ep}(p,-p')-\chi_{F,2\ep}(p,-p')]\\
  &\hspace{90pt}-\frac{\sin[b(p-p')]}{2\pi(p+p')}\big[F(|p|)-F(|p'|)\big]\bigg]
  \ov{\wh{\psi}(p)}\wh{\psi'}(p')\,dpdp'\,.
 \end{split}\label{mdn-4}
\end{equation}
Finally, we note that
\begin{equation}\label{mdn-free}
    (\psi,F(h_z)\psi')
    =\int F(|r|)\ov{\wh{\psi}(r)}\wh{\psi'}(r)\,dr\,.
\end{equation}
Setting now expressions (\ref{mdn-1} -- \ref{mdn-4}) into
(\ref{mdn-F}) and subtracting (\ref{mdn-free}) we find
\begin{equation}
    (\psi,[F(\hb)-F(h_z)]\psi')=\int
    \ov{\wh{\psi}(p)}\,K^F_a(p,p')\,
    \wh{\psi'}(p')\,dpdp'\,,\label{mdn-FF}
\end{equation}
and as a consequence obtain Eq.\,(\ref{mdn-GFF}), where the
kernels are given by
\begin{equation}
 \begin{split}\label{mdn-kerF}\\
    &K^F_a(p,p')=\cos[b(p-p')][k_F+k_{F,\ep}](p,p')\\
    &\hspace{140pt}+\cos[b(p+p')][k_{F,\ep}-k_{F,2\ep}](p,p')\\
    &=\cos[b(p-p')][k_F+k^*_{F,\ep}](p,p')
    +\cos[b(p+p')][k^*_{F,\ep}-k^*_{F,2\ep}](p,p')\\
    &\hspace{120pt}+2\sin(bp)\sin(bp')k^0_{F,\ep}(p,p')\,,
 \end{split}
\end{equation}
with
\begin{equation}\label{mdn-k}
 \begin{split}
    &k_F(p,p')=\frac{1}{2\pi^2}
    \Big[\chi_F(p,p')-\da\chi_F(p,-p')\Big]\,,\\
    &k_{F,\ep}(p,p')=\frac{1}{2\pi^2}
    \Big[\chi_{F,\ep}(p,p')-\da\chi_{F,\ep}(p,-p')\Big]\,,
 \end{split}
\end{equation}
and similarly for $k^0_{F,\ep}$, $k^*_{F,\ep}$, and
\begin{equation}\label{mdn-kerFG}
       K^{F,G}_a(p,p')=G(|p|)K^F_a(p,p')G(|p'|)\,
\end{equation}
(all denotations as at the beginning of Appendix \ref{pw}). Using
the second formula in (\ref{mdn-kerF}) and looking at the
estimates (\ref{pw-echi1} -- \ref{pw-echi4}) one finds that both
functions $K_a^{F,G}(p,p')$ and $\sqrt{|p|}K_a^{F,G}(p,p')$ are
square-integrable. In more detail: square integrability on the
square $(p,p')\in\<-1,1\rangle^2$ follows rather immediately from
the estimates (\ref{pw-echi1} -- \ref{pw-echi3}) (the term
containing $k^0_{F,\vep}$ has to be analysed jointly with the sine
functions -- see (\ref{mdn-kerF})). For the square integrability
on $\mathbb{R}^2\setminus\<-1,1\rangle^2$ one uses
(\ref{pw-echi4}) and (\ref{mdn-fgb}); it is obvious that in this
case the integration of the estimates can be harmlessly extended
to $\mathbb{R}^2$, and then by symmetry narrowed to
$\<0,\infty)^2$. In~this way one finds
\begin{equation}\label{mdn-r2}
\begin{aligned}
 &\int_{\mathbb{R}^2\setminus\<-1,1\rangle^2}
 [1+|p|]\big|K_a^{F,G}(p,p')\big|^2dpdp'\\
 &\hspace*{50pt}\leq\con \int_{0\leq p\leq p'}\frac{1+p+p'}
 {(p+1)^{2(\alpha+\gamma)}(p'+1)^{2(\alpha+1)}}dpdp'<\infty\,,
\end{aligned}
\end{equation}
where for the integration over $p\geq p'\geq0$ the variables have
been swapped, and the condition $4\alpha+2\gamma>1$ has been taken
into account (cf.\ (\ref{mdn-fgb})). The formula (\ref{mdn-ce})
for the Casimir energy is now easily obtained (cf.\
Eq.\,(\ref{en-energy})). Finally, to obtain the scaling behaviour
it is easily checked that $K^{F_\mu,G_\mu}_a(p,p')=K^{F,G}_{\mu
a}(p/\mu,p'/\mu)$, which upon substitution into (\ref{mdn-ce})
gives the desired result.

We end this section with remarks on the relation of the present
work to that of~\cite{her}. The models considered in the cited
work are obtained by choosing the Dirichlet boundary conditions
for $\hb$ and setting $G\equiv1$. One may thus ask whether it is
possible to set $G\equiv1$ also in the case of Neumann conditions,
which would alow to bring the analysis to simpler terms of
\cite{her}. It turns out that it is not, and this posed the major
difficulty in extending the program to these boundary conditions.
The estimates (\ref{pw-echi1} -- \ref{pw-echi4}) guarantee that
$K_a^F(p,p')$ is square-integrable, but are insufficient for the
square-integrability of $\sqrt{|p|}K_a^F(p,p')$ (this is easily
seen by putting $\alpha=0$ in (\ref{mdn-r2})). However, if one
takes into account that the functions $\chi$ enter the kernel only
through the combinations (\ref{mdn-k}), then it turns out that the
terms in $\chi$'s decaying most slowly for $|p|,|p'|\to\infty$
subtract in the case of Dirichlet, but add in the case of Neumann
conditions, and the function $\sqrt{|p|}K_a^F(p,p')$ is
square-integrable in the former, but not in the latter case.

\setcounter{equation}{0}

\section{Asymptotic expansion of the Casimir energy}\label{asy}

In this section we show how to expand the Casimir energy of the
class of models defined in the previous section into inverse
powers of the distance of plates $a$. We shall assume that $F$ and
$G$ are as for strict boundary conditions in some neighbourhood of
zero, $F(p)=p$ and $G(p)=1$ for small $p$. This assumption is
somewhat stronger then the most economic one, which demands only
that a few derivatives of those functions in $p=0$ agree with
those of the above special functions. We choose the simpler
version to shorten the proofs. Also, we assume the
differentiability to arbitrary order, which spares us tedious
track-keeping. We note that to take $\mu$ arbitrarily large in
(\ref{mdn-scaled}) and (\ref{mdn-scaling}) physically means to
approach strict boundary conditions. The asymptotic expansion
combined with the scaling determines the behaviour of the Casimir
energy in that limit.

Below we prove the following.
\begin{itemize}
\item[(Asym)]Let the functions $F,G\in\C^\infty\big(\<0,\infty)\big)$
satisfy the conditions
\begin{equation}
 \begin{aligned}
 &0\leq F(p)\leq\con\,,\quad F(p)\leq p\,,\quad 0\leq G(p)\leq1\,,\\
 &|F^{(n)}(p)|\leq\con(n)\,(p+1)^{-n-\gamma}\,,\quad
 n=1,2,\ldots\,,\\
 &|G^{(k)}(p)|\leq\con(k)\,(p+1)^{-k-\alpha}\,,\qquad
 k=0,1,2,\ldots\\
 &\text{for some}\ \alpha,\gamma\in(0,1)\ \text{and such that}\
 2(\alpha+\gamma)>1\,,
 \end{aligned}\qquad
 \label{asy-bfgd}
\end{equation}
and let moreover $F(p)=p$ and $G(p)=1$ in some neighbourhood of
zero. Then
\begin{equation}\label{asy-asy}
    \vep_a=\vep_\infty+(\tfrac{1+\delta}{2})\frac{c}{a^2}
    -\frac{\pi^2}{1440a^3}+\vep_{4,a}\,,
\end{equation}
where $\vep_\infty$ and $c$ are model ($F$ and $G$)-dependent
constants and
\begin{equation}\label{asy-este}
    |\vep_{4,a}|\leq\frac{\con}{a^4}\,,\quad
    \Big|\frac{d\vep_{4,a}}{da}\Big|\leq\frac{\con}{a^5}\,.
\end{equation}
The limit value $\vep_\infty$ is twice the Casimir energy $\vep_0$
for the configuration of one single plate,
\begin{equation}\label{asy-infsin}
 \vep_\infty=2\vep_0\,.
\end{equation}
 The scaling limit
$\mu\to\infty$ yields
\begin{equation}\label{asy-sl}
    \lim_{\mu\to\infty}\Big(\vep^\mu_a-\vep_\infty\mu^3
    -(\tfrac{1+\delta}{2})\frac{c}{a^2}\mu\Big)=-\frac{\pi^2}{1440a^3}\,.
\end{equation}
Constants $\vep_\infty$ and $c$ are given in (\ref{asy-einf}),
(\ref{asy-c1}), (\ref{asy-c2}) and (\ref{asy-c}) below.
\end{itemize}

\noindent As already mentioned above we do not strive to optimize
assumptions on functions $F$ and $G$, which could be substantially
weakened, but rather want to simplify the proof. The assumptions
on derivatives essentially mean that the functions do not have
oscillatory behaviour; the stronger (then in (Mod)) bound on
parameters $\alpha$ and $\gamma$ will be needed in our expansion
procedure.

The usual interpretation of formula (\ref{asy-sl}) in the
Dirichlet case would be that apart from an ``unimportant
constant'' the limit of strict boundary conditions gives the
original Casimir expression. However, this is not true for the
Neumann conditions. Moreover, the constant $\vep_\infty$ does have
physical meaning: this is the energy needed for the creation of
the configuration of the field surrounding infinitely separated
plates. Relation (\ref{asy-infsin}) shows that configurations
surrounding each of the plates become independent in the large
separation limit (at least as far as energy is concerned). We also
note, although we do not present explicit calculations, that in
the sharp boundary limit also the particle number per area
diverges.

One could ask how generic are the lessons drawn from (Asym). We
think that the result whose generality is least prone to doubt is
the nonexistence of sharp boundaries limit. We support this by two
observations. First, we recall that the free quantum field and the
field in presence of sharp boundaries are even not described by
the same algebra of observables, not to mention their
representations (this was pointed out in \cite{AH}). Second, we
learn that quantities which are responsible for comparability of
representations and for backreaction, $n_a$ and $\vep_a$, become
infinite. Now, note that these are \emph{positive} quantities,
having physical interpretation. These facts indicate very strongly
to our conclusion. Similar views on the unphysical nature of sharp
boundary conditions may be found in literature, but we are not
aware of a rigorous discussion based on the investigation of the
algebraic structure of the theory and modification of dynamics
(see also discussion and bibliography in \cite{AH}).

Also, we note that our calculation offers a rigorous derivation of
the original Casimir term $-\pi^2/1440a^2$ in the backreaction
energy and confirms its universality. On the other hand the
derivation shows, that it cannot stand alone: the energy is the
expectation value in certain state of a positive operator, so it
must be positive. Even if one adds a positive, constant
($a$-independent) term to the Casimir term one does not get a
positive expression for each separation value $a$.

The appearance of a term quadratic in $1/a$ in the Neumann case
stands probably more open to debate. However, we want to point out
that our rigorous analysis should rather be viewed as giving rise
to a \emph{converse} problem: are there any models for the
dynamics $h_{za}$, for which the quadratic term would be absent?;
what would distinguish them? Within our class of models there is
no general reason for $c$ to vanish, and at this point we are not
aware whether the condition $c=0$ has any solutions in that class.
In fact, a partial result to the opposite in certain circumstances
may be shown (see the end of this section), for brevity we
simplify assumptions on functions $F$ and $G$ to a more special
class then needed. Suppose that in addition to the assumptions of
(Asym) we have $G(p)=f(p-r)$ for $p\geq0$, where $r$ is a positive
parameter and $f(x)$ is a smoothed step function equal to $1$ for
$x\leq -\kappa$ and to $0$ for $x\geq\kappa$, with some
$\kappa>0$. Then for sufficiently large $r$ constant $c$ is
positive. Note that increasing $r$ moves us towards better
approximation of sharp boundaries. Further discussion of the
physical meaning of our results will be found at the end of
Section \ref{el}.

We shall assume in the proof of (Asym) that $F(p)=p$ and $G(p)=1$
for $p\in\<0,1\rangle$. This does not restrict generality: each
case of functions satisfying the assumptions of (Asym) may be
brought to this more restrictive case by
rescaling~(\ref{mdn-scaled}).

\subsection{Asymptotic value of the Casimir energy}

We first consider the Casimir energy for the $z$-motion dynamics
of modified single plate. Let $h^B_{z0}$ be determined by
$(h^B_{z0})^2=-(\p^2/(\p z)^2)_B$ with Dirichlet or Neumann
conditions on the plane $z=0$, and define $h_{z0}$ as in
Eq.\,(\ref{mdn-mod}). The discussion of the last section
simplifies greatly in that case and shows that $h_{z0}-h_z$ is
again an integral operator in the momentum representation, with
the kernel
\begin{equation}\label{asy-ker0}
    K^{F,G}_0(p,p')= G(|p|)k_F(p,p')G(|p'|)\,,
\end{equation}
with $k_F$ as defined by (\ref{mdn-k}). The Casimir energy per
unit surface $\vep_0$ for this operator (formula (\ref{mdn-ce})
with this new kernel) is the energy one needs to create the
configuration of the field in the lowest stationary state of the
modified Hamiltonian, as discussed in \cite{AH}.

We now consider the limit value $\vep_\infty$ in our model of
modified parallel plates. For $a\to\infty$, i.e.\ $\ep\to0$, one
has for $p\neq0$ the point-wise limit
$\dsp\lim_{\ep\to0}\La_{F,\ep}(p)=\La_F(p)$. This is easily shown
with the use of formula (\ref{sum-EM}) and the estimates
(\ref{pw-intla1}) and (\ref{pw-intla2}). Then for $p,p'\neq0$,
$p\neq p'$, there is
$\dsp\lim_{\ep\to0}\chi^*_{F,\ep}(p,p')=\chi_F(p,p')$, and
$\dsp\lim_{\ep\to0}\chi^0_{F,\ep}(p,p')=0$. At~the same time
referring back to the estimations discussed towards the end of
Section \ref{mdn} one realizes that the functions
$G(|p|)[\chi^*_{F,\ep}\!-\!\chi_F](p,p')G(|p'|)$,
\mbox{$\sqrt{|p|}G(|p|)[\chi^*_{F,\ep}-\chi_F](p,p')G(|p'|)$},
$\sin(bp)\sin(bp')G(|p|)\chi^0_{F,\ep}(p,p')G(|p'|)$ and
$\sqrt{|p|}\sin(bp)\sin(bp')G(|p|)\chi^0_{F,\ep}(p,p')G(|p'|)$
remain bounded in modulus by square-integrable functions
independent of $\ep$ (when it is small). Hence for $\ep\to0$ we
have
\begin{gather*}
 \|G(|p|)[\chi^*_{F,\ep}-\chi_F](p,p')G(|p'|)\|_{L^2}\to0\,,\\
 \big\|\sqrt{|p|}\,G(|p|)[\chi^*_{F,\ep}-\chi_F](p,p')
 G(|p'|)\big\|_{L^2}\to0\,,\\
 \|\sin(bp)\sin(bp')G(|p|)\chi^0_{F,\ep}(p,p')G(|p'|)\|_{L^2}\to0\,,\\
 \big\|\sqrt{|p|}\sin(bp)\sin(bp')G(|p|)\chi^0_{F,\ep}(p,p')G(|p'|)\big\|_{L^2}\to0\,,
\end{gather*}
so for the purpose of calculating the limit $a\to\infty$ one can
replace the kernel $K^{F,G}_a(p,p')$ with
$\cos[b(p-p')]\,2K^{F,G}_0(p,p')$ in the formula (\ref{mdn-ce}).
Using now the identities:
\begin{align}
 &\cos^2[b(p-p')]=\frac{1}{2}\Big\{1+\cos[2b(p-p')]\Big\}\,,
 \label{asy-cos2}\\
 \begin{split}
 &\cos[b(p-p')]\cos[b(p'-p'')]\cos[b(p''-p)]\\
 &\hspace{50pt}=\frac{1}{4}\Big\{1+\cos[2b(p-p')]+\cos[2b(p'-p'')]
 +\cos[2b(p''-p)]\Big\}
 \end{split}\label{asy-cos3}
\end{align}
together with the Riemann-Lebesgue lemma we find
\begin{multline}\label{asy-einf}
    \vep_\infty=2\frac{1}{24\pi}
    \bigg\{\int 3|p|[K^{F,G}_0(p,p')]^2\,dp\,dp'\\
    +\int K^{F,G}_0(p,p')K^{F,G}_0(p',p'')K^{F,G}_0(p'',p)
    \,dp\,dp'\,dp''\bigg\}=2\vep_0\,.
\end{multline}

\subsection{Decomposition $\boldsymbol{\vep_a=\vep_a^{\mathrm{(i)}}
+\vep_a^{\mathrm{(ii)}}}$}

We now turn to the derivation of further leading terms of the
asymptotic expansion of $\vep_a$ for $a\to\infty$. It will be
convenient to denote
\begin{equation}\label{asy-D}
    D^F_a=F(\hb)-F(h_z)
\end{equation}
and
\begin{equation}\label{asy-g}
    g(u)=1-G^2(u)\,.
\end{equation}
We note that the operators $D^F_a$ and $\sqrt{h_z}G(h_z)D^F_a$ are
HS -- this is shown by methods used towards the end of Section
\ref{mdn} for the estimation of kernels (and here for the second
one of these operators we need the strengthened condition on
$\alpha$ and~$\gamma$, see (\ref{asy-bfgd})). Using this notation
and manipulating the operators under the trace sign we obtain
\begin{equation}
 \begin{split}
 &\Tr(h_{za}-h_z)^3=\Tr\big[\big(D^F_a\big)^3(1-3g(h_z))\big]\\
 &\hspace{35pt}+3\Tr\big[\big(D^F_a\big)^2g(h_z)D^F_ag(h_z)\big]
 -\Tr\big[D^F_ag(h_z)D^F_ag(h_z)D^F_ag(h_z)\big]\,,
 \end{split}
\end{equation}
\begin{equation}
 \begin{split}
 &\Tr\big[(h_{za}-h_z)h_z(h_{za}-h_z)\big]=
 \Tr\big[\sqrt{h_z}G(h_z)\big(D^F_a\big)^2G(h_z)\sqrt{h_z}\big]\\
 &\hspace{180pt}-\Tr\big[D^F_aG^2(h_z)h_zD^F_ag(h_z)\big]\,.
 \end{split}
\end{equation}
We set these expressions into the formula for the Casimir energy,
in addition we add and subtract the term
$\Tr\big[D^F_aF(h_z)D^F_a(1-3g(h_z))\big]$, and write the result
in the form
\begin{gather}
 \vep_a=\vep_a^{\mathrm{(i)}}+\vep_a^{\mathrm{(ii)}}\,,\label{asy-esplit}\\
 \begin{split}
 &\vep_a^{\mathrm{(i)}}=\frac{1}{24\pi}
 \Tr\Big\{\big[\big(D^F_a\big)^3+D^F_aF(h_z)D^F_a\big](1-3g(h_z))\\
 &\hspace{70pt}+3\sqrt{h_z}G(h_z)\big(D^F_a\big)^2G(h_z)\sqrt{h_z}
 -\lp D^F_a\rp^2F(h_z)\Big\}\,,
 \end{split}\label{asy-e11}\\
 \begin{split}
 &\vep_a^{\mathrm{(ii)}}
 =\frac{1}{24\pi}\Tr\Big\{3D^F_a\big[F(h_z)-G^2(h_z)h_z\big]D^F_ag(h_z)\\
 &\hspace{50pt}+3\big(D^F_a\big)^2g(h_z)D^F_ag(h_z)
 -D^F_ag(h_z)D^F_ag(h_z)D^F_ag(h_z)\Big\}\,.
 \end{split}\label{asy-e21}
\end{gather}
This decomposition of $\vep_a$ is purely technical: different
techniques will be now employed for the calculation of the
asymptotic expansion in each of the two cases
$\vep_a^{\mathrm{(i)}}$ and $\vep_a^{\mathrm{(ii)}}$. However, it
could help to observe that if it was possible to put $G\equiv1$
(as can, in fact, be done in the Dirichlet case -- see remarks
towards the end of Section~\ref{mdn}), then the second term in
(\ref{asy-esplit}) would vanish (as then $g\equiv0$).  More
comments on this point will be found below.

\subsection{Integral representation of $\boldsymbol{\vep_a^{\mathrm{(i)}}}$}

By elementary manipulations one finds the identities
\begin{align}
    &\big(D^F_a\big)^2=D^{F^2}_a-F(h_z)D^F_a-D^F_aF(h_z)\,,\label{asy-D2}\\
    \begin{split}
    &\big(D^F_a\big)^3+D^F_aF(h_z)D^F_a\\
    &\hspace{40pt}=D^{F^3}_a-F(h_z)D^{F^2}_a
    -D^{F^2}_aF(h_z)+F(h_z)D^F_aF(h_z)\,.
    \end{split}\label{asy-D3}
\end{align}
Operators $D^{F^2}_a$ and $D^{F^3}_a$ are HS, with the kernels in
momentum representation $K^{F^2}_a,K^{F^3}_a\in L^2(\mR^2)$. The
above two identities, when written in terms of kernels, take the
form
\begin{equation}
    \int K^F_a(p,q)K^F_a(q,p')\,dq=
    K^{F^2}_a(p,p')-\big[F(|p|)+F(|p'|)\big]\,K^F_a(p,p')\,,\label{asy-D2ker}
\end{equation}
\begin{equation}
    \begin{split}
    &\int K^F_a(p,q)K^F_a(q,q')K^F_a(q',p')\,dq\,dq'+
    \int K^F_a(p,q)F(|q|)K^F_a(q,p')\,dq\\
    &=K^{F^3}_a(p,p')-\big[F(|p|)+F(|p'|)\big]\,K^{F^2}_a(p,p')
    +F(|p|)K^F_a(p,p')F(|p'|)\,.
    \end{split}\label{asy-D3ker}
\end{equation}
These identities hold in the $L^2$-sense. However, as both sides
are continuous functions of $(p,p')$ for $p,p'\neq0$, they are
also valid there point-wise, in particular also for $p=p'$. Using
this fact for the calculation of (\ref{asy-e11}) one finds
\begin{equation}\label{asy-e12}
 \begin{split}
    &\vep_a^{\mathrm{(i)}}=\frac{1}{24\pi}\int\bigg\{\big[1-3g(|p|)\big]K^{F^3}_a(p,p)\\
    &\hspace{60pt}+3\big[G^2(|p|)|p|-F(|p|)+2F(|p|)g(|p|)\big]K^{F^2}_a(p,p)\\
    &\hspace{120pt}+3G^2(|p|)F(|p|)\big[F(|p|)-2|p|\big]K^F_a(p,p)
    \bigg\}\,dp\,.
 \end{split}
\end{equation}
We note that this expression engages only the \emph{diagonal}
values $K_a(p,p)$ of the kernels. In fact, as part
$\vep_a^{\mathrm{(i)}}$ is concerned, one could bypass the
calculation of the kernels and use a simpler technique employed in
Ref.\,\cite{her}. But once we have the kernels (needed for the
calculation of $\vep_a^{\mathrm{(ii)}}$), it is convenient to use
them also here.

Looking at eq.\,(\ref{mdn-kerF}) we see that
\begin{equation}\label{asy-kfpp}
    K^F_a(p,p)=[k_F+k_{F,\ep}](p,p)
    +\cos(ap)[k_{F,\ep}-k_{F,2\ep}](p,p)\,.
\end{equation}
Using the definitions (\ref{pw-La}) and (\ref{pw-chi}) we find
$\chi_F(p,p)=\La^{(1)}_F(p)$ and $\chi_F(p,-p)=\La_F(p)/p$ and
then recalling (\ref{mdn-k}) we have
\begin{equation}\label{asy-pp}
    k_F(p,p)
    =\frac{1}{4\pi^2}\int
    \Big\{\Big(\frac{1}{p-q}+\frac{\da}{p+q}\Big)^2
    \big[F(|q|)-F(|p|)\big]-\frac{2|p|F^{(1)}(|p|)}{(q-p)(q+p)}\Big\}\,dq\,.
\end{equation}
In a similar way one finds
\begin{equation}\label{asy-ppep}
    k_{F,\vep}(p,p)
    =\frac{\vep}{4\pi^2}\sum_{k\in\mathbb{Z}}
    \Big\{\Big(\frac{1}{p-k\vep}+\frac{\da}{p+k\vep}\Big)^2
    \big[F(|k\vep|)-F(|p|)\big]-\frac{2|p|F^{(1)}(|p|)}{(k\vep-p)(k\vep+p)}\Big\}\,.
\end{equation}
Also, analogous formulas for $F^2$ or $F^3$ replacing $F$ in
(\ref{asy-kfpp} -- \ref{asy-ppep}) are valid. Setting all these
formulas into Eq.\,(\ref{asy-e12}) one finds that the terms
proportional to $F^{(1)}(|p|)$ cancel out and one is left with
\begin{equation}\label{asy-e13}
 \begin{split}
    &\vep_a^{\mathrm{(i)}}=\frac{1}{24\pi^3}\int \rho(q,p)\,dq\,dp\\
    &\hspace{20pt}+\frac{1}{24\pi^3}\int\Big\{[1+\cos(ap)]\,
    \ep\sum_{k\in\mZ}\rho(k\ep,p)
    -\cos(ap)\,2\ep\sum_{k\in\mZ}\rho(2k\ep,p)\Big\}\,dp\,,
 \end{split}
\end{equation}
where
\begin{equation}\label{asy-rho}
 \begin{split}
    &\rho(q,p)=\la_F^2(q,p)
    \Big[\tfrac{1+\delta}{2}\,p^2+\tfrac{1-\delta}{2}\,q^2\Big]\\
    &\hspace{100pt}\times\Big[F(|q|)[1-3g(|p|)]-F(|p|)+3G^2(|p|)|p|\Big]\,.
 \end{split}
\end{equation}
All integrals and sums are absolutely convergent. Both here and in
further expressions below involving $\rho$ and its derivatives one
uses the estimates found in Appendix \ref{pw} to prove
convergence. We do not go into easy but tedious details.

\subsection{Expansion of $\boldsymbol{\vep_a^{\mathrm{(i)}}}$}

As $\rho(q,p)$ is even in each of the variables, one can reduce
integration and summation to nonnegative values. We evaluate the
sum $\dsp\ep\sum_{k\in\mZ}\rho(k\ep,p)$ for fixed $p>0$ with the
use of formula (\ref{sum-EM}) for $n=4$:
\begin{equation}\label{asy-sume}
 \begin{split}
    &\ep\sum_{k\in\mZ}\rho(k\ep,p)
    =2\int_0^\infty\rho(q,p)\,dq
    -\ep^2B_2\rho^{(1,0)}(0,p)
    -2\ep^4R_{\rho,4,\ep}(p)\,,
 \end{split}
\end{equation}
where
\begin{equation}\label{asy-R}
    R_{\rho,4,\ep}(p)=\frac{1}{4!}\int_0^\infty\rho^{(4,0)}(q,p)b_4(q/\ep)\,dq
\end{equation}
and $\rho^{(1,0)}(0,p)$ is the limit value of $\rho^{(1,0)}(q,p)$
for $q\searrow0$. To obtain the other sum in the integrand of
Eq.\,(\ref{asy-e13}) one needs only to replace $\ep$ by $2\ep$.
The integrability of $|R_{\rho,4,\ep}(p)|$ at infinity is easily
shown with the use of estimates to be found in Appendix C. On the
other hand, these estimates do not guarantee the integrability for
small $p$. However, this property will be obvious from what
follows below. Setting the above expansion into
Eq.\,(\ref{asy-e13}) we have
\begin{equation}\label{asy-e14}
 \begin{split}
    \vep_a^{\mathrm{(i)}}=&\frac{1}{3\pi^3}\int_0^\infty\rho(q,p)\,dq\,dp
    -\frac{\ep^2}{72\pi^3}
    \int_0^\infty\rho^{(1,0)}(0,p)[1-3\cos(ap)]\,dp\\
    +&\frac{1}{6\pi^3}\int_0^\infty\Big\{-[1+\cos(ap)]\,
    \ep^4R_{\rho,4,\ep}(p)
    +\cos(ap)\,(2\ep)^4R_{\rho,4,2\ep}(p)\Big\}\,dp\,.
 \end{split}
\end{equation}
The first term on the r.h.\ side is $a$-independent and it
contributes to $\vep_\infty$. To~evaluate the second term we find
explicitly
\begin{equation}\label{asy-rho10}
    \rho^{(1,0)}(0,p)=-3(\tfrac{1+\da}{2})\eta(p)\,,\quad
    \eta(p)\equiv G^2(p)\frac{F(p)[2p-F(p)]}{p^2}>0\,.
\end{equation}
Using this one finds
\begin{equation}\label{asy-sec}
 \begin{split}
    &-\frac{\ep^2}{72\pi^3}
    \int_0^\infty\rho^{(1,0)}(0,p)[1-3\cos(ap)]\,dp\\
    &\hspace{20pt}=\big(\tfrac{1+\da}{2}\big)
    \bigg\{\frac{1}{24\pi a^2}\int_0^\infty \eta(p)\,dp
    +\frac{1}{8\pi a^3}\int_0^\infty
    \eta^{(1)}(p)\sin(ap)\,dp\bigg\}\,,
 \end{split}
\end{equation}
where integration by parts in the second term has been performed.
From (\ref{asy-rho10}) we find that $\eta(p)=1$ for
$p\in\<0,1\rangle$, so $\eta^{(1)}(p)=0$ in this interval. It is
easily seen by standard Fourier transform properties that the
second integral in the last formula and its derivative with
respect to $a$ vanish faster than any inverse power of $a$ for
$a\to\infty$ (by repeated integration by parts and use of
estimates (\ref{asy-bfgd})). The first integral contributes only
to the constant $c$ in (\ref{asy-asy}), and yields
\begin{equation}\label{asy-c1}
    c^\mathrm{(i)}=\frac{1}{24\pi}
    \int_0^\infty G^2(p)\frac{F(p)[2p-F(p)]}{p^2}\,dp\,.
\end{equation}

Next, to evaluate the second line on the r.h.\ side of
Eq.\,(\ref{asy-e14}) we split
\mbox{$R_{\rho,4,\ep}=R_{\rho_\mathrm{r},4,\ep}+R_{\rho_\mathrm{s},4,\ep}$},
where
\begin{equation}\label{asy-rhosp}
    \rho=\rho_\mathrm{r}+\rho_\mathrm{s}\,,\qquad
    \rho_\mathrm{s}(q,p)=\frac{p^3}{(p+q)^2}+\da\frac{p^2}{p+q}\,.
\end{equation}
For $q,p\in\<0,1\rangle$ we have
$\rho(q,p)=\rho_\mathrm{s}(q,p)+(1-\da)q/2$, so $\rho_\mathrm{r}$
is regular in a neighbourhood of zero. Thus using (\ref{asy-R}) we
find the estimate
\begin{equation}\label{asy-estreg}
 \begin{split}
    &\bigg|\frac{1}{6\pi^3}\int_0^\infty\Big\{-[1+\cos(ap)]\,
    \ep^4R_{\rho_\mathrm{r},4,\ep}(p)
    +\cos(ap)\,(2\ep)^4R_{\rho_\mathrm{r},4,2\ep}(p)\Big\}\,dp\bigg|\\
    &\hspace{150pt}\leq\frac{\con}{a^4}
    \int_0^\infty|\rho^{(4,0)}_\mathrm{r}(q,p)|\,dq\,dp
    <\infty\,.
 \end{split}
\end{equation}
For the integrability of $|\rho^{(4,0)}_\mathrm{r}(q,p)|$ one uses
the assumption (\ref{asy-bfgd}), relation (\ref{asy-rhosp}) and
the results of Appendix C. Furthermore, using the relations
\begin{gather}\label{asy-rder}
    \begin{aligned}
    \int_0^\infty\frac{\p\cos(ap)}{\p a}R_{\rho_\mathrm{r},4,\ep}(p)\,dp
    &=\frac{1}{a}\int_0^\infty\frac{\p\cos(ap)}{\p p}pR_{\rho_\mathrm{r},4,\ep}(p)\,dp\\
    &\hspace*{-2em}=-\frac{1}{a}\int_0^\infty\cos(ap)
    [pR^{(1)}_{\rho_\mathrm{r},4,\ep}(p)+
    R_{\rho_\mathrm{r},4,\ep}(p)]\,dp\,,
    \end{aligned}\\
    R^{(1)}_{\rho_\mathrm{r},4,\ep}(p)
    =\frac{1}{4!}\int_0^\infty\rho_\mathrm{r}^{(4,1)}(q,p)b_4(q/\ep)\,dq\,,
\end{gather}
\begin{equation}
  \begin{split}
    &\frac{\p}{\p\ep}[\ep^4R_{\rho_\mathrm{r},4,\ep}(p)]
    =\frac{\p}{\p\ep}\frac{\ep^5}{4!}
    \int_0^\infty\rho_\mathrm{r}^{(4,0)}(t\ep,p)b_4(t)\,dt\\
    &\hspace{100pt}=5\ep^3R_{\rho_\mathrm{r},4,\ep}(p)
    +\frac{\ep^3}{4!}\int_0^\infty q\rho_\mathrm{r}^{(5,0)}(q,p)
    \,b_4(q/\ep)\,dq\,,
  \end{split}
\end{equation}
one also finds
\begin{equation}\label{asy-estregda}
 \begin{split}
    &\bigg|\frac{d}{da}\ \frac{1}{6\pi^3}\int_0^\infty\Big\{-[1+\cos(ap)]\,
    \ep^4R_{\rho_\mathrm{r},4,\ep}(p)
    +\cos(ap)\,(2\ep)^4R_{\rho_\mathrm{r},4,2\ep}(p)\Big\}\,dp\bigg|\\
    &\hspace{10pt}\leq\frac{\con}{a^5}
    \int_0^\infty\Big[|\rho^{(4,0)}_\mathrm{r}(q,p)|
    +p|\rho^{(4,1)}_\mathrm{r}(q,p)|
    +q|\rho^{(5,0)}_\mathrm{r}(q,p)|\Big]\,dq\,dp
    <\infty\,.
 \end{split}
\end{equation}
Integrability of $p|\rho^{(4,1)}_\mathrm{r}(q,p)|$ and of
$q|\rho^{(5,0)}_\mathrm{r}(q,p)|$ is shown as for
$|\rho^{(4,0)}_\mathrm{r}(q,p)|$ above.

The second term resulting from the split of $\rho$, that involving
$R_{\rho_\mathrm{s},4,\ep}$, can be explicitly evaluated. We have
\begin{gather}
    \ep^4R_{\rho_\mathrm{s},4,\ep}(p)
    =\ep^2R_{\rho_\mathrm{s},4,1}(p/\ep)\,,\label{asy-sing1}\\
     R_{\rho_\mathrm{s},4,1}(t)=(-t^3\p_t+\da\,t^2)
    \frac{1}{4!}\int_0^\infty
    \Big(\p_u^4\frac{1}{t+u}\Big)b_4(u)\,du\,.\label{asy-sing2}
\end{gather}
With the use of formula (\ref{sum-EMg}) with $n=4$ and $\ep=1$,
applied to the function $f(u)=1/(t+u)$ (for fixed $t$) we find
\begin{equation}\label{asy-psi}
 \begin{split}
    \frac{1}{4!}\int_0^\infty
    \Big(\p_u^4\frac{1}{t+u}\Big)b_n(u)\,du
    &=\lim_{N\to\infty}\bigg\{\int_0^N\frac{du}{t+u}
    -\sum_{k=0}^N\frac{1}{t+k}\bigg\}
    +\frac{1}{2t}+\frac{B_2}{2t^2}\\
    &=\psi(t)-\log t+\frac{1}{2t}
    +\frac{B_2}{2t^2}\equiv \frac{w_2(t)}{t^2}\,,
 \end{split}
\end{equation}
where we used formula (\ref{psi-rep1}) and notation introduced in
(\ref{psi-w}), and then
\begin{equation}\label{asy-sing3}
    R_{\rho_\mathrm{s},4,1}(t)
    =-tw^{(1)}_2(t)+(2+\da)w_2(t)\,.
\end{equation}
Setting this into (\ref{asy-sing1}) and using the identities
(\ref{psi-F2}) and (\ref{psi-F2prim}) we find
\begin{equation}\label{asy-sing4}
 \begin{split}
    &\frac{1}{6\pi^3}\int_0^\infty\Big\{-[1+\cos(ap)]\,
    \ep^4R_{\rho_\mathrm{s},4,\ep}(p)
    +\cos(ap)\,(2\ep)^4R_{\rho_\mathrm{s},4,2\ep}(p)\Big\}\,dp\\
    &\hspace{20pt}
    =\frac{1}{6a^3}\int[1+\cos(\pi t)-8\cos(2\pi t)]\,
    [tw^{(1)}(t)-(2+\da)w(t)]\,dt\\
    &\hspace{120pt}=-\frac{1}{16\pi^2}\zeta(4)
    =-\frac{\pi^2}{1440\,a^3}\,.
 \end{split}
\end{equation}
To arrive at the last line in this formula one has to take into
account on the r.h.\ side of (\ref{psi-F2prim}) that
\[
 \dsp\sum_{k=1}^\infty\frac{1}{(k+\frac{1}{2})^s}
 =(2^s-1)\sum_{k=1}^\infty\frac{1}{k^s}-2^s\,.
\]

\subsection{Integral representation and expansion of
$\boldsymbol{\vep_a^{\mathrm{(ii)}}}$}

We now turn to $\vep_a^{\mathrm{(ii)}}$ -- the second term in
(\ref{asy-esplit}). We use identity (\ref{asy-D2}) in formula
(\ref{asy-e21}) and calculate the trace in the momentum
representation, which yields
\begin{equation}\label{asy-e22}
 \begin{split}
    \vep_a^{\mathrm{(ii)}}&=\frac{1}{8\pi}\int
    K^F_a(p,p')K^{F^2}_a(p',p)g(|p|)g(|p'|)\,dp\,dp'\\
    &-\frac{1}{8\pi}\int
    \big[K^F_a(p,p')\big]^2g(|p|)\Big\{G^2(|p'|)|p'|-F(|p'|)
    +2F(|p'|)g(|p'|)\Big\}\,dp\,dp'\\
    &-\frac{1}{24\pi}\int K^F_a(p,p')K^F_a(p',p'')
    K^F_a(p'',p)g(|p|)g(|p'|)g(|p''|)\,dp\,dp'\,dp''\,.
 \end{split}
\end{equation}
This expression looks much more complicated than the integral
formula (\ref{asy-e12}) for $\vep_a^{\mathrm{(i)}}$ (multiple
integrals and full kernels $K_a(p,p')$). However, a substantial
simplification of the asymptotic expansion will be due to the
fact, that the integrations here stay away from the singularities
of the kernels $K_a(p,p')$, which occur in the neighbourhood of
$p=0$ or $p'=0$ (as for $u\in\<0,1\rangle$ there is $G(u)=1$,
$g(u)=0$ and $F(u)=u$ according to our assumptions) and the
integrands are, consequently, infinitely differentiable. Namely,
products of kernels $K_a$ contain products of cosine functions,
which may be expressed as linear combinations of other cosine
functions and, possibly, unity (as in Eqs.\,(\ref{asy-cos2}),
(\ref{asy-cos3})). Those terms in the integrals above which
contain one of the resulting cosines yield functions of $a$ which
vanish faster than any inverse power of $a$ for $a\to\infty$. This
is shown by standard Fourier methods: integrate by parts and use
decay properties of the integrand and its derivatives -- estimates
(\ref{asy-bfgd}) and (\ref{pw-echi5}); in the process the decay
rate of the integrand increases. One also shows that the
$a$-derivative of those terms has similar decay properties.
Indeed, the derivative when acting on trigonometric functions does
not change the decay; if it acts on a sum of the form $\ep\sum
f(k\ep)$ it yields $-(1/a)\ep\sum[f(k\ep)+k\ep f^{(1)}(k\ep)]$.
These facts are used for the proof of the last statement, but we
do not go into further straightforward details.

Thus it is sufficient to consider only those terms which take a
constant from the product of cosines. One finds easily that the
only products of cosines in the above integrals which do give
constants are those in which cosines with ``+'' sign between the
variables do not occur, or occur twice. However, we shall see
below that these parts of the kernels $K$ which multiply the
cosines with the ``+'' sign are of order $\ep^2$ at least, so the
quadratic terms of that type do not contribute to the orders
$\leq3$ in $\ep$ in the energy, and to the orders $\leq4$ in the
force. Thus we shall be left only with the products
(\ref{asy-cos2}) and (\ref{asy-cos3}).

To expand the kernels $K_a$ we use the definition (\ref{pw-Laep})
and the formula (\ref{sum-EM}) to get
\begin{align}\label{asy-La}
    \La_{F,\ep}(p)&=\La_F(p)+\frac{\ep^2}{6p}
    -\ep^4\La_{F,4,\ep}(p)\,,\\
    \La_{F^2,\ep}(p)&=\La_{F^2}(p)
    -\ep^4\La_{F^2,4,\ep}(p)\,,
\end{align}
where
\begin{equation}\label{asy-La4}
    \La_{F^i,4,\ep}(p)
    =\frac{1}{4!}2p\int_0^\infty\la_{F^i}^{(4,0)}(q,p)
    b_4(q/\ep)\,dq\,,\quad i=1,2\,.
\end{equation}
Setting these expressions into (\ref{pw-chiep}) and (\ref{mdn-k})
one finds
\begin{equation}\label{asy-kasy}
 \begin{aligned}
    k_{F,\ep}(p,p')&=k_F(p,p')
    -\frac{1+\da}{12\pi^2}\frac{\ep^2}{pp'}-\ep^4k_{F,4,\ep}(p,p')\,,\\
    k_{F^2,\ep}(p,p')&=k_{F^2}(p,p')-\ep^4k_{F^2,4,\ep}(p,p')\,,
 \end{aligned}
\end{equation}
where for $i=1,2$:
\begin{equation}\label{asy-k4}
    k_{F^i,4,\ep}(p,p')=\frac{1}{2\pi^2}
    \bigg[\frac{\La_{F^i,4,\ep}(p)-\La_{F^i,4,\ep}(p')}{p-p'}
    -\da\frac{\La_{F^i,4,\ep}(p)+\La_{F^i,4,\ep}(p')}{p+p'}\bigg]
\end{equation}
Note that $|p|,|p'|\geq1$ for our present purposes, so there are
no singularities. Then with the use of bounds similar to those in
(\ref{pw-echi5}) one shows that the terms containing $k_{F,4,\ep}$
indeed do not contribute in the given orders. Thus omitting the
last term in both formulas (\ref{asy-kasy}) and taking into
account the result of the discussion following
Eq.\,(\ref{asy-e22}) we find that disregarding terms higher than
$(1/a)^3$ in $\vep_a^{\mathrm{(ii)}}$ we can make the following
replacements in (\ref{asy-e22}) (recall the formula
(\ref{mdn-kerF})):
\begin{equation}
 \begin{aligned}
 &K_a^F(p,p')K_a^{F^2}(p'p)\rightarrow
 \frac{1}{2}\Big[2k_F(p,p')-\frac{1+\da}{12\pi^2}\frac{\ep^2}{pp'}\Big]
 2k_{F^2}(p',p)\,,\\
 &[K_a^F(p,p')]^2\rightarrow
 \frac{1}{2}\Big[2k_F(p,p')-\frac{1+\da}{12\pi^2}\frac{\ep^2}{pp'}\Big]^2\,,\\
 &K^F_a(p,p')K^F_a(p',p'')K^F_a(p'',p)\rightarrow\\
 &\hspace{.0em}\rightarrow\frac{1}{4}
 \Big[2k_F(p,p')-\frac{1+\da}{12\pi^2}\frac{\ep^2}{pp'}\Big]
 \Big[2k_F(p',p'')-\frac{1+\da}{12\pi^2}\frac{\ep^2}{p'p''}\Big]
 \Big[2k_F(p'',p)-\frac{1+\da}{12\pi^2}\frac{\ep^2}{p''p}\Big]\,.
 \end{aligned}
\end{equation}
In this way we obtain
\begin{equation}\label{asy-e23}
  \vep^\mathrm{(ii)}_a=
  \big(\tfrac{1+\delta}{2}\big)\frac{c^\mathrm{(ii)}}{a^2}
  +\text{terms independent of}\ a+ O(a^{-4})\,,
\end{equation}
where
\begin{equation}\label{asy-c2}
    \begin{split}
    &c^\mathrm{(ii)}=
 -\frac{1}{12\pi}\int_0^\infty k_{F^2}(p,p')\frac{g(p)g(p')}{pp'}\,dp\,dp'\\
   &\hspace{30pt}+\frac{1}{6\pi}\int_0^\infty k_F(p,p')\frac{g(p)}{p}
   \bigg[G^2(p')-\frac{F(p')}{p'}+2\frac{F(p')}{p'}g(p')\bigg]\,dp\,dp'\\
   &\hspace{30pt}+\frac{1}{6\pi}\int_0^\infty k_F(p,p')k_F(p,p'')
   \frac{g(p)g(p')g(p'')}{p'p''}\,dp\,dp'\,dp''\,.
   \end{split}
\end{equation}
The constant $c$ in (\ref{asy-asy}) is now
\begin{equation}\label{asy-c}
    c=c^\mathrm{(i)}+c^\mathrm{(ii)}\,.
\end{equation}
This ends the proof of Eqs.\,(\ref{asy-asy}) and (\ref{asy-este}).
Equation (\ref{asy-sl}) is their simple consequence.

Finally, we want to prove the statement on positivity of $c$ in
circumstances described towards the end of the discussion
following the formulation of (Asym). We note that the term
$c^{\mathrm{(i)}}$ obtained in (\ref{asy-c1}) is then a positive,
increasing function of $r$ tending to the limit value
\begin{equation}
 \lim_{r\to\infty}c^\mathrm{(i)}(r)=\frac{1}{24\pi}
 \int_0^\infty \frac{F(p)[2p-F(p)]}{p^2}dp\,.
\end{equation}
At the same time one shows with the use of the present assumptions
on $G(p)$ and estimates on functions $k$ that
$|c^\mathrm{(ii)}(r)|\leq\con/r^\gamma$, so
\begin{equation}
 \lim_{r\to\infty}c^\mathrm{(ii)}(r)=0\,,
\end{equation}
which ends the proof.

\setcounter{equation}{0}

\section{Electromagnetic field}\label{el}

When dealing with a scalar field we have first formulated the free
field model in terms of the value of the field, and its
time-derivative, on a Cauchy hyperplane. Then we formulated the
dynamics in presence of the external influence in terms of the
same variables, and took over for the unperturbed energy of the
field itself (at a given time) the old expression built with the
use of them. In this way a simple model could be obtained in which
the ``sources'' of the field were supplied by the field variables
themselves. A similar procedure for the electromagnetic field
needs some caution. The main problem lies in the fact that the
dynamics of the field is constrained, and one of the constraints
depends on sources (Gauss law). The evolution is governed by a
first order equation and one could think of the electric and
magnetic parts of the field as the analogues of the scalar field
variables at a given time, but then the initial data for the
electric part are differently constrained in free and interacting
case. Therefore one first has to solve constraints and identify
independent field variables. The modification of the dynamics of
these variables must then produce an imitation of the perfect
conductor boundary conditions of the electromagnetic field. To
achieve this, one has to adjust the choice of the independent
fields to the geometry of the problem. We shall first describe our
choice in the case of the classical free field restricted to the
region $\Omega\Car\mR$, where $\Omega$ is a compact convex region
in the $x$-$y$ plane, and $\mR$ is the $z$-axis. The following
informal discussion serves to motivate the precise formulation of
the model, which starts after Eq.\,(\ref{el-en}).

For a vector $\vec{A}$ we denote by $\vec{A}_\bot$ its part
perpendicular to the $z$-axis, and by $A$ its $z$-component. By
${}^*\!\vec{A}_\bot$ we denote the dual vector in the $x$-$y$
plane, that is in a Cartesian basis ${}^*\!A^1=A^2$,
${}^*\!A^2=-A^1$. The free Maxwell equations are then
\begin{align}
  \vec{\nabla}_\bot\cdot\vec{B}_\bot+\p_zB&=0\,,&
  \vec{\nabla}_\bot\cdot\vec{E}_\bot+\p_zE&=0\,,\label{el-m1}\\
  \vec{\nabla}_\bot\cdot{}^*\vec{B}_\bot-\p_tE&=0\,,&
  \vec{\nabla}_\bot\cdot{}^*\vec{E}_\bot+\p_tB&=0\,,\label{el-m2}\\
  \p_z\vec{B}_\bot-\p_t{}^*\vec{E}_\bot&=\vec{\nabla}_\bot B\,,&
  \p_z\vec{E}_\bot+\p_t{}^*\vec{B}_\bot+&=\vec{\nabla}_\bot E\,.\label{el-m3}
\end{align}
Locally, of course, the wave equation follows for each of the
components, in particular
\begin{equation}\label{el-wa}
    [\p_t^2-\p_z^2-\Delta_\bot]B=0\,,\quad
    [\p_t^2-\p_z^2-\Delta_\bot]E=0\,.
\end{equation}
Global extension to the whole region demands some boundary
conditions for $\Delta_\bot$ in each case. We assume that the
Dirichlet extension for the electric case and the Neumann
extension for the magnetic case have been chosen. Assume moreover
that $\int_\Omega B\,dx\,dy=0$, that is $B$ is orthogonal to
constants in $L^2(\Omega)$. Then one can represent these fields as
\begin{equation}\label{el-EB}
    B=-\Delta_\bot \Psi_m\,,\quad
    E=-\Delta_\bot \Psi_e\,,
\end{equation}
where $\Psi_m$ is assumed to be orthogonal to constants, and the
fields $\Psi_m$ and $\Psi_e$ are thus uniquely determined by $B$
and $E$. Equations (\ref{el-wa}) can be now expressed as
\begin{equation}\label{el-wV}
    [\p_t^2-\p_z^2-\Delta_\bot]\Psi_m=0\,,\quad
    [\p_t^2-\p_z^2-\Delta_\bot]\Psi_e=0\,,
\end{equation}
with appropriate boundary conditions. It is now easy to see that
setting
\begin{equation}\label{el-per}
    \vec{B}_\bot=\vec{\nabla}_\bot\p_z \Psi_m
    +{}^*\vec{\nabla}_\bot\p_t \Psi_e\,,\quad
    \vec{E}_\bot=\vec{\nabla}_\bot\p_z \Psi_e
    -{}^*\vec{\nabla}_\bot\p_t \Psi_m
\end{equation}
one solves the complete set of Maxwell equations. The transversal
fields satisfy boundary conditions
\begin{equation}\label{el-trb}
    \vec{n}\cdot\vec{B}_\bot=0\,,\quad
    {}^*\vec{n}\cdot\vec{E}_\bot=0\,,
\end{equation}
where $\vec{n}$ is a vector in the $x$-$y$ plane, orthogonal to
$\p\Omega$. Let us now add to $B$ the omitted part $B'$
independent of $\vec{x}_\bot$, and to $\vec{B}_\bot$ and
$\vec{E}_\bot$ additional fields $\vec{B'}_\bot$ and
$\vec{E'}_\bot$ respectively in order to investigate uniqueness.
We want to keep boundary conditions (\ref{el-trb}). One finds then
by integrating the first equation in (\ref{el-m1}) and the second
equation in (\ref{el-m2}) over $\Omega$ that $B'$ is in fact a
constant, which we exclude. Then it is easy to show that the only
solution for the remaining fields $\vec{B'}_\bot$ and
$\vec{E'}_\bot$ is zero.

Our boundary conditions are those of a perfect conductor, but this
is not necessary in all its details. For instance, one could
interchange the roles of the electric and magnetic fields. More
generally, the boundary conditions should eliminate
$\vec{x}_\bot$-independent fields. However, for definiteness we
keep our choice.

The electromagnetic potential $(\vp,\vec{A}_\bot,A)$ producing
fields (\ref{el-EB}), (\ref{el-per}) may be chosen as
\begin{equation}\label{el-pot}
    \vp=-\p_z\Psi_e\,,\quad
    \vec{A}_\bot={}^*\vec{\nabla}_\bot \Psi_m\,,\quad
    A=\p_t \Psi_e\,.
\end{equation}
However, the model is defined in terms of gauge-independent
quantities and the potential is a purely auxiliary field.

Using the boundary conditions one finds the total energy of the
field in terms of $\Psi_e$ and $\Psi_m$:
\begin{equation}\label{el-en}
 \begin{split}
    &\frac{1}{8\pi}
    \int[\vec{E}^2+\vec{B}^2](t,\vec{x})\,d^3x\\
    &=\frac{1}{8\pi}\sum_{s=e,m}\int[\p_t\Psi_s(-\Delta_\bot)\p_t\Psi_s+
    \p_z\Psi_s(-\Delta_\bot)\p_z\Psi_s
    +(-\Delta_\bot \Psi_s)^2](t,\vec{x})\,d^3x\,.
 \end{split}
\end{equation}

This brings us to the following formulation of the classical
dynamics of our system. We use for this formulation variables
$V_e$ and $V_m$ introduced below, which should be thought of as
supplying Cauchy data for fields $\Psi_e$ and $\Psi_m$
respectively, i.e.\ \mbox{$V_s=(\Psi_s,\partial_t\Psi_s)|_{t=0}$},
$s=e,m$. The electromagnetic field is derived from these
variables.

Let $L^2_\mR(\Omega)_\bot$ denote the subspace of
$L^2_\mR(\Omega)$ orthogonal to constants,
\mbox{$h^D_\bot=\sqrt{-\Delta^D_\bot}$} and
$h^N_\bot=\sqrt{-\Delta^N_\bot}$, where $\Delta^D_\bot$ is the
Dirichlet selfadjoint Laplacian on $L^2_\mR(\Omega)$, and
$\Delta^N_\bot$ is the Neumann selfadjoint Laplacian on
$L^2_\mR(\Omega)_\bot$. Generally, let $C$ be a selfadjoint
operator on a Hilbert space $\mathcal{H}$. By using a spectral
representation of $C$ it is then easy to see the following: if $C$
is positive and has a bounded inverse, then
$\mathcal{H}_C\equiv\D(C)$ is a Hilbert space with respect to the
scalar product $(\vp,\psi)_C=(C\vp,C\psi)$. Moreover, the
restriction of $C$ to $\D(C^2)$ is a selfadjoint, positive
operator in $\mathcal{H}_C$, with spectrum equal to that of $C$.
We apply this statements to the operators $h^D_\bot$ and
$h^N_\bot$ defined above, and denote
\begin{gather}\label{el-sp1}
    \R_\bot^e=\D(h^D_\bot)\,,\qquad
    (\vp,\psi)_{\bot e}=\frac{1}{4\pi}(h^D_\bot\vp,h^D_\bot\psi)\,,\\
    \R_\bot^m=\D(h^N_\bot)\,,\qquad
    (\vp,\psi)_{\bot m}=\frac{1}{4\pi}(h^N_\bot\vp,h^N_\bot\psi)\,\\
    \R_z=L^2_\mR(\mR)\,,\quad
    \R^e=\R^e_\bot\otimes\R_z\,,\quad
    \R^m=\R^m_\bot\otimes\R_z\,,
\end{gather}
and the scalar product in $\R^s$ is denoted by $(.\,,.)_s$.
Moreover, we denote by $h_\bot^e$ and $h_\bot^m$ the selfadjoint
operators in $\R_\bot^e$ and $\R_\bot^m$ determined in the above
way by $h_\bot^D$ and $h_\bot^N$ respectively, and then set
\begin{equation}\label{el-hem}
    h^e=\sqrt{(h_\bot^e\otimes\id)^2+(\id\otimes h_z)^2}\,,\qquad
    h^m=\sqrt{(h_\bot^m\otimes\id)^2+(\id\otimes h_z)^2}\,,
\end{equation}
where $h_z=\sqrt{-\p_z^2}$. The space of our model is now
\begin{equation}\label{el-sym}
    \mathcal{L}=\mathcal{L}^m\oplus\mathcal{L}^e\,,\quad
    \mathcal{L}^s=\D(h^s)\oplus\R^s\,,\quad s=m,e\,.
\end{equation}
We write $V_s=v_s\oplus u_s\in\mathcal{L}^s$, $s=m,e$. The spaces
$\mathcal{L}^s$ are equipped with symplectic forms
\begin{equation}\label{el-sf}
    \sigma^s(V_s,V'_s)=(v_s',u_s)_s-(v_s,u'_s)_s\,.
\end{equation}
The Hamiltonian of the system is then
\begin{equation}\label{el-clham}
    \mathcal{H}(V_m,V_e)=\mathcal{H}^m(V_m)+\mathcal{H}^e(V_e)\,,\quad
    \mathcal{H}^s(V_s)=\frac{1}{2}
    \big[(u_s,u_s)_s+(h^sv_s,h^sv_s)_s\big]\,,
\end{equation}
and the evolution in each of the spaces $\mathcal{L}^s$ is
independently determined by $\mathcal{H}^s$ as a symplectic
transformation, as discussed in Sec.\,I-3. One can now show that
for sufficiently regular fields this reproduces the evolution
equations (\ref{el-wV}) and the total energy of the field
(\ref{el-en}).

It is now evident that our classical system is described by the
direct sum of two systems of the type discussed in Section
\ref{sc}. If we put
\mbox{$\Omega=\<-L_x,L_x\rangle\Car\<-L_y,L_y\rangle$}, then
$h_\bot^e$ and $h_\bot^m$ have the spectrum of the type (D) and
(N) respectively. The quantization of each independent part of the
system follows the same lines as before. The algebra of quantum
variables of the entire system is then the $C^*$-tensor product of
the two Weyl algebras. The modification of the dynamics brought
about by the modified boundary conditions is implemented by the
replacement of (\ref{el-hem}) by
\begin{equation}\label{el-hemb}
    h_a^e=\sqrt{(h_\bot^e\otimes\id)^2+(\id\otimes h_{za}^e)^2}\,,\qquad
    h_a^m=\sqrt{(h_\bot^m\otimes\id)^2+(\id\otimes h_{za}^m)^2}\,,
\end{equation}
where $h_{za}^s$, $s=e,m$, are constructed as in (\ref{mdn-mod}),
with $h_{za}^B$ determined by Neumann conditions for the case
$s=e$ and by the Dirichlet conditions for the case $s=m$ (which
imitates perfect conductor conditions). The representations needed
for the discussion of the Casimir effect are constructed as tensor
products of the two subsystems, and then the energy observable is
the sum of the energies of the subsystems. Thus, in particular,
the Casimir energy is the sum of the Casimir energies of the two
subsystems, which proves our claim formulated in Introduction.

Our strategy in the above analysis was to formulate the classical
dynamics in terms of unconstrained and gauge-independent
variables, and then quantize. For our purposes we did not need the
quantum version of the electromagnetic potential or field itself.
However, for the completeness we briefly sketch their formulation.
We obtain the electromagnetic field from the potential, and for
the latter we keep the quantum version of the formulas
(\ref{el-pot}). Recall that the fields independent of
$\vec{x}_\bot$ are to be excluded. Moreover, we do not expect the
formulas for potential (or electromagnetic field) to extend to the
boundary of $\Omega$. Thus we take for our test function space
$\Delta_\bot\D_\Omega$, where $\D_\Omega$ denotes the space of
infinitely differentiable functions with compact support contained
in $\Omega\Car\mR$. Denote by $v_s^\mathrm{q}$ and
$u_s^\mathrm{q}$ heuristic quantized versions of $v_s$ and $u_s$
respectively. Then motivated by (\ref{el-pot}) we put for
$f\in\D_\Omega$
\begin{equation}\label{el-qpoth}
  \begin{split}
    &\vp(\Delta_\bot f)=\int v_e^\mathrm{q}\Delta_\bot \p_zf\,d^3x=
    -4\pi(v_e^\mathrm{q},\p_zf)_e\,,\\
    &A(\Delta_\bot f)=\int u_e^\mathrm{q}\Delta_\bot f\,d^3x=
    -4\pi(u_e^\mathrm{q},f)_e\,,\\
    &\vec{A}_\bot(\Delta_\bot f)=-\int v_m^\mathrm{q}
    {}^*\vec{\nabla}_\bot\Delta_\bot f\,d^3x=
    4\pi(v_m^\mathrm{q},{}^*\vec{\nabla}_\bot f)_m\,.
  \end{split}
\end{equation}
Now, the precise meaning of $v_s^\mathrm{q}$ and $u_s^\mathrm{q}$
is given by
\begin{equation}\label{el-quant}
    \Phi_s(V_s)=(v_s^\mathrm{q},u_s)_s+(u_s^\mathrm{q},v_s)_s\,,
\end{equation}
where $\Phi_s(V_s)$ are quantum fields as described in Section
I-3. Therefore the precise meaning of (\ref{el-qpoth}) is
\begin{gather}\label{el-qpot}
    \vp(\Delta_\bot f)=-4\pi\Phi_e(0,\p_z f)\,,\qquad
    A(\Delta_\bot f)=-4\pi\Phi_e(f,0)\,,\\
    \vec{A}_\bot(\Delta_\bot f)
    =4\pi\Phi_m(0,{}^*\vec{\nabla}_\bot f)\,.
\end{gather}
The electromagnetic field is then easily found:
\begin{equation}\label{el-qfi}
  \begin{split}
    &E(\Delta_\bot f)
    =4\pi\Phi_e(0,[\Delta_\bot-(h_{za}^e)^2+h_z^2] f)\,,\\
    &B(\Delta_\bot f)=4\pi\Phi_m(0,\Delta_\bot f)\,,\\
    &\vec{E}_\bot(\Delta_\bot f)
    =-4\pi\Phi_e(0,\vec{\nabla}_\bot\p_z f)-
    4\pi\Phi_m({}^*\vec{\nabla}_\bot f,0)\,,\\
    &\vec{B}_\bot(\Delta_\bot f)
    =4\pi\Phi_e({}^*\vec{\nabla}_\bot f,0)
    -4\pi\Phi_m(0,\vec{\nabla}_\bot\p_z f)\,.
  \end{split}
\end{equation}
The formula for $E$ depends on dynamics; for the free field
$h_{za}^e$ should be replaced by $h_z$, and then $E(\Delta_\bot f)
=\Phi_e(0,\Delta_\bot f)$. Using the commutation relations
\begin{equation}\label{el-cr}
    [\Phi_s(V_s),\Phi_s(V'_s)]=i\sigma^s(V_s,V'_s)\id
\end{equation}
one finds equal time commutators
\begin{equation}\label{el-crem}
 \begin{split}
    &[E^i_\bot(\Delta_\bot f),B^j_\bot(\Delta_\bot g)]
    =-4\pi i\ep^{ij}\int(\Delta_\bot f)
    \p_z(\Delta_\bot g)\,d^3x\,,\\
    &[B(\Delta_\bot f),\vec{E}_\bot(\Delta_\bot g)]
    =4\pi i\int (\Delta_\bot f)
    {}^*\vec{\nabla}_\bot(\Delta_\bot g)\,d^3x\,,\\
    &[E(\Delta_\bot f),\vec{B}_\bot(\Delta_\bot g)]
    =-4\pi i\int (\Delta_\bot f)
    {}^*\vec{\nabla}_\bot(\Delta_\bot g)\,d^3x\\
    &\hspace{120pt}+4\pi i\int \Big([(h_{za}^e)^2-h_z^2]f\Big)\,
    {}^*\vec{\nabla}_\bot(\Delta_\bot g)\,d^3x\,\,,
 \end{split}
\end{equation}
where $\ep^{ij}$ is the antisymmetric symbol with $\ep^{12}=1$,
and all other commutators vanish. For the free field the last term
in the third relation vanishes and the commutators reproduce the
usual quantization scheme. Also, the free Maxwell equations then
hold for fields (\ref{el-qfi}). In the presence of the modified
plates the commutators cannot remain unchanged, as this would
violate the constraints; the last term in the third commutator
above takes care of that. The modification is nonlocal -- this
reflects the nonlocality of our model ($(h_{za}^e)^2$ is
nonlocal). The pair of homogeneous Maxwell equations is still
satisfied for fields (\ref{el-qfi}) -- by definition, which may be
regarded as gauge invariance --  while the r.h.\ sides of the
other two give the sources. We note, moreover, that the
identification of the full electromagnetic field in the
interacting case is subject to some arbitrariness. The sources are
linear functionals of the same fields, so one could shift some
part of the electromagnetic fields to the r.h.\ sides of
inhomogeneous equations to contribute to sources (but provided the
homogeneous equations are conserved under the operation). Our
choice of the interpretation is the most simple one.

We end with a summary of the physical meaning of our results. We
believe that our analysis places the Casimir effect firmly within
standard quantum theory. On the other hand it also shows why less
conscious traditional quantum field formulations suffer from
difficulties. The models analysed here are well defined and are
free from the usual anomalies. The Casimir energy can be
rigorously calculated for them as the expectation value of the
positive free field energy operator in the ground state enforced
by the environment (in accordance with the analysis presented in
\cite{AH}).

In the case of the electromagnetic field our results may have
direct physical application. It follows from the discussion
following (Asym) in Section \ref{asy} that one should not expect a
universal law for the total Casimir force between modified
parallel plates. The original Casimir expression $-\pi^2/720a^3$
constitutes the third order term of the expansion of the energy in
inverse powers of separation $a$, and cannot give a dominant
contribution to the force for all values of this parameter. In
particular, energy is positive, so the term is dominated by other
nonconstant contributions for small~$a$. On the other end of the
range of $a$ we have found that in our models typical fall-of of
Casimir energy is governed by a term of order $1/a^2$. This
prediction may be of more special character, but we believe that
if it is indeed, then disappearance of such a term in any other
models needs further explanation. What we find in our setting is
that for a wide range of models the coefficient at this term is
positive, so the force at sufficiently large $a$ becomes
repulsive.

Eventually, these predictions must be checked against experiment.
However, although we have witnessed impressive progress in
precision measurements of the Casimir force in recent years, and
the existence of the force is now beyond doubt, its detailed form
seems to us less certain. Due to experimental difficulties a
measurement of the force between parallel plates has been reported
only relatively recently \cite{bcor}. The results seem to confirm
the original Casimir formula. However, we think it is to early to
accept this as ``the whole truth'' on this force. First of all,
the measurements span a limited range of the distance parameter
$a$. As explained above we have very strong reasons to believe
that the Casimir formula cannot hold for $a$ tending to zero. The
behaviour of the force for large $a$ could also bring deviations.
Second, the data analysis seems to be oriented at testing the
coefficient at the $1/a^4$ term of the measured force. To a large
extent it involves peeling off other influences, while keeping the
$\sim 1/a^4$ formula for the backreaction force; the best fit then
confirms the original Casimir coefficient at this term. When seen
from that angle the result does not contradict our predictions. We
think that more extensive experimental tests of the functional
form of the force are needed.

\setcounter{equation}{0}

\section*{Appendices}
\appendix

\section{Sums and integrals}\label{sum}

In this appendix we gather a handful of approximation formulas
connecting sums with integrals. Some of these results are
well-known, but we give them the form needed in the main text. In
particular, the Euler-Maclaurin expansion is usually formulated
for analytical functions only, with no estimates on the rest.
\begin{itemize}
\item[$\mr{(i)_A}$] Let $f:(0,\infty)\mapsto \<0,\infty)$ be a
non-increasing, continuous function. Then
\begin{gather}
 \lim_{\ep\to 0}\ep\sum_{k=1}^\infty f(k\ep)=\int_0^\infty
 f(u)\,du\,,\label{sum-si1}\\
 \lim_{\ep_x,\ep_y\to0}\ep_x\ep_y\sum_{k,\hspace{1pt}l=1}^\infty f(k^2\ep_x^2+l^2\ep_y^2)
 =\frac{\pi}{4}\int_0^\infty f(u)\,du\,,\label{sum-si2}
\end{gather}
where the equalities hold both for finite and infinite integral on
the right hand side.  If in addition
\begin{align}
 &\int_0^1f(u)\,du<\infty &&\text{then}\qquad
 \lim_{u\to0} uf(u)=0\,,\label{sum-li1}\\
 &\int_1^\infty f(u)\,du<\infty &&\text{then}\qquad
 \lim_{u\to\infty} uf(u)=0\,,\label{sum-li2}\\
 &\int_0^\infty f(u)\,du<\infty &&\text{then}\qquad
 \lim_{\ep\to0}\ep^2\sum_{k=1}^\infty f(k^2\ep^2)=0\,.\label{sum-li3}
\end{align}
\item[$\mr{(ii)_A}$] Euler-Maclaurin expansion\\
Let $f$ be a complex function in $\C^n\big(\<0,\infty)\big)$ for
some $n\in\mN$ and such that
\begin{gather}
 \int_0^\infty |f^{(n)}(u)|\,du<\infty\,,\\
 \lim_{u\to\infty}f^{(m)}(u)=0\quad\text{for}\quad
 m=0,1,3,5,\ldots,\leq n-2\,.
\end{gather}
Then the following identity holds
\end{itemize}
\begin{equation}\label{sum-EMg}
 \begin{split}
  &\lim_{N\to\infty}\bigg\{\frac{\ep}{2} f(0)
  +\ep\sum_{k=1}^{N}f(k\ep)
  -\int_0^{N\ep}f(u)\,du\bigg\}\\
  &\hspace{150pt}=-\sum_{m=2}^{n-1}
  \ep^m\frac{B_m}{m!}f^{(m-1)}(0)-\ep^nR_{f,n,\ep}\,,
 \end{split}
\end{equation}
\indent where
\begin{equation}
 R_{f,n,\ep}=\frac{1}{n!}\int_0^\infty
 f^{(n)}(u)b_n(u/\ep)\,du\,,\label{sum-R}
\end{equation}
\begin{equation}
 \begin{split}
 b_n(k+s)=b_n(s)\equiv\, &B_n(1-s)-\tfrac{1}{2}B_n(0)
 -\tfrac{1}{2}B_n(1)\\
 &\hspace{0pt}\text{for}\ s\in\<0,1)\,,\ k\in\mN\,,
 \end{split}\label{sum-bn}
\end{equation}
\begin{equation}
 |R_{f,n,\ep}|\leq c_n\int_0^\infty|f^{(n)}(s)|\,du\,.\label{sum-Rb}
\end{equation}
\begin{itemize}
\item[] Here $B_n(.)$ are Bernoulli polynomials, $B_n=B_n(0)$ are
Bernoulli constants~\cite{gr}, and $\dsp
c_n=(n!)^{-1}\max_{s\in\<0,1\rangle}|B_n(s)-B_n|$.

In particular, if in addition the integral $\dsp\int_0^\infty
f(u)\,du$ converges  then also the sum $\dsp\ep\sum_{k=1}^\infty
f(k\ep)$ does (neither needs to converge absolutely), and in that
case the identity can be written as
\end{itemize}
\begin{equation}\label{sum-EM}
 \frac{\ep}{2} f(0)+\ep\sum_{k=1}^\infty f(k\ep)=\int_0^\infty f(u)\,du
 -\sum_{m=2}^{n-1}\ep^m\frac{B_m}{m!}f^{(m-1)}(0)-\ep^nR_{f,n,\ep}\,.
\end{equation}

\subsection{Proof of $\boldsymbol{\mr{(i)_A}}$}

The first identity (\ref{sum-si1}) follows immediately from
Riemann approximations for a positive, non-increasing, continuous
function
\begin{equation}\label{sum-ineq}
 \int_\ep^\infty f(u)\,du\leq
 \ep\sum_{k=1}^\infty f(k\ep)\leq
 \int_0^\infty f(u)\,du\,.
\end{equation}
The application of these inequalities to $f(\alpha^2+\beta^2)$ as
a function of $\alpha$ and $\beta$ leads to
\begin{multline}\label{sum-ineq'}
 \int_{\ep_x}^\infty\int_{\ep_y}^\infty
 f(\alpha^2+\beta^2)\,d\alpha\, d\beta\leq
 \ep_x\ep_y\sum_{k,\hspace{1pt}l=1}^\infty f(k^2\ep_x^2+l^2\ep_y^2)\\
 \leq \int_0^\infty\int_0^\infty
 f(\alpha^2+\beta^2)\,d\alpha\, d\beta
 =\frac{\pi}{4}\int_0^\infty f(u)\,du\,,
\end{multline}
which proves the identity (\ref{sum-si2}).

The function $f$ is non-increasing, so $\dsp uf(u)\leq
2\int_{u/2}^u f(v)\,dv$, which implies the properties
(\ref{sum-li1}) and (\ref{sum-li2}). By the first of these limits
the first term in the sum in Eq.\,(\ref{sum-li3}) tends to zero.
For the rest of the sum, for $\ep<1$,  we have
\begin{multline}\label{sum-ineq2}
 \ep^2\sum_{k=2}^\infty f(k^2\ep^2)\leq
 \ep\int_\ep^\infty f(\alpha^2)\,d\alpha
 =\frac{\ep}{2}\int_{\ep^2}^\infty \frac{f(u)}{\sqrt{u}}\,du\\
 \leq \frac{1}{2}\int_{\ep^2}^\ep f(u)\,du
 +\frac{\sqrt{\ep}}{2}\int_\ep^\infty f(u)\,du\,,
\end{multline}
where the first inequality follows by the Riemann approximation of
the integral. This is sufficient to conclude that
Eq.\,(\ref{sum-li3}) holds.

\subsection{Proof of $\boldsymbol{\mr{(ii)_A}}$}

One first notes that for $n\in\mN$ the following identity holds
\begin{equation}\label{sum-indid}
\begin{split}
 &\frac{\ep}{2}f(0)+\ep\sum_{k=1}^{N-1}f(k\ep)+\frac{\ep}{2}f(N\ep)
 -\int_0^{N\ep}f(s)\,ds\\
 &\hspace{20pt}=-\sum_{m=2}^{n-1}\ep^m\frac{B_m}{m!}[f^{(m-1)}(0)-f^{(m-1)}(N\ep)]\\
 &\hspace{33pt}-\frac{\ep^n}{n!}\sum_{k=0}^{N-1}\int_0^\ep f^{(n)}(k\ep+s)
 [B_n(1-(s/\ep))-\tfrac{1}{2}B_n(0)-\tfrac{1}{2}B_n(1)]\,ds
\end{split}
\end{equation}
This is shown by induction with respect to $n$, with the use of
integration by parts in the integrals in the second sum on the
r.h.\ side (remember that
\mbox{$\frac{d}{du}B_n(u)=nB_{n-1}(u)$}). The change of
integration variable $u=k\ep+s$ in each of the integrals in this
sum puts the second line on the r.h.\ side into the form
\[
 -\frac{\ep^n}{n!}\int_0^{N\ep}f^{(n)}(u)b_n(u/\ep)du\,.
\]
Functions $b_n$ are measurable and bounded, so the estimate
(\ref{sum-Rb}) follows. Therefore, if the assumptions are
satisfied then the r.h.\ side of (\ref{sum-indid}), and the term
$(\ep/2)f(N\ep)$ on the l.h. side converge for $N\to\infty$. All
the statements of the thesis are now readily seen.

\setcounter{equation}{0}

\section{Hilbert-Schmidt properties of operators
$\boldsymbol{\Delta_a(u)}$}\label{hs}

In this appendix $h_z$ and $h_{za}$ are arbitrary selfadjoint,
positive operators with domains $\D(h_z)$ and $\D(h_{za})$
respectively, and operators $h(u)$ and $h_a(u)$ are defined by
Eqs.\,(\ref{sc-hu}) and (\ref{sc-hua}) respectively (we keep this
notation to make application of the results in the main text
obvious, but we do not need any further general restrictions). Let
$f$ and $g$ be real (or complex) measurable functions on
$\<0,\infty)$ such that $f(u)$, $g(u)$, $uf(u)$ and $ug(u)$ are
all bounded on the domain. Then the expressions
\begin{equation}
\begin{split}
 &f(h_{za})\Delta^{p,q}_a(u)g(h_z)\,,\quad \text{where
 formally}\\
  &\Delta^{p,q}_a(u)=h^{-p}_a(u)\Delta_a(u)h^{-q}(u)\,,\qquad
  \Delta_a(u)=h_a(u)-h(u)\,,
 \end{split}
\end{equation}
define bounded operators for all $p,q\geq0$ and all
$u\in(0,\infty)$ ($u\in\<0,\infty)$ if\linebreak $p=q=0$),
although in general expressions $\Delta^{p,q}_a(u)$ need not make
sense by themselves. We denote by $P_B$ and $P_{aB}$ the spectral
projectors determined by $h_z$ and $h_{za}$ respectively,
projecting onto the set $B$ indicated in the subscripts. In
particular, the intervals in the symbols $P_{\<\beta_1,\beta_2)}$
and $P_{a\<\gamma_1,\gamma_2)}$ below are chosen inside
$\<0,\infty)$ (positivity of operators). Symbol $\|A\|_\HS$ will
be used for the Hilbert-Schmidt norm of a HS operator $A$, i.e.\
$\|A\|_\HS=\sqrt{\Tr[A^*A}]$.

We shall show in this appendix that under these assumptions the
following identities and implications hold. The principal results,
which are needed in the main text, are contained in $\mr{(iv)_B}$
and $\mr{(v)_B}$ below.
\begin{itemize}
\item[$\mr{(i)_B}$] For each $u>0$ and $v\geq0$ the following
integral representation holds in the uniform sense
\end{itemize}
\begin{equation}\label{hs-fg}
 f(h_{za})\Delta_a(u)g(h_z)=\frac{1}{\pi}\int_u^\infty
 \frac{f(h_{za})}{h_{za}^2+t}\big[h_a(v)\Delta_a(v)+\Delta_a(v)h(v)\big]
 \frac{g(h_z)}{h_z^2+t}\sqrt{t-u}\,dt\,.
\end{equation}
\begin{itemize}
\item[$\mr{(ii)_B}$] If
 $P_{a\<\gamma_1,\gamma_2)}\Delta^{p,q}_a(u)
 P_{\<\beta_1,\beta_2)}$
is a HS operator for a given $u=v\geq0$, then it is also a HS
operator for $u\in(0,\infty)$, and
\end{itemize}
\begin{equation}\label{hs-hsp}
\begin{split}
 \|P_{a\<\gamma_1,\gamma_2)}\Delta^{p,q}_a(u)
 P_{\<\beta_1,\beta_2)}\|_\HS&\leq
 \|P_{a\<\gamma_1,\gamma_2)}\Delta^{p,q}_a(v)
 P_{\<\beta_1,\beta_2)}\|_\HS\\
 \times\frac{\sqrt{\gamma_2^2+v}+\sqrt{\beta_2^2+v}}
 {\sqrt{\gamma_1^2+u}+\sqrt{\beta_1^2+u}}&\times
 \begin{cases}1&\text{for}\ u>v\,,\\
 \big(\frac{v}{u}\big)^{(p+q)/2}&\text{for}\ u\leq v\,.
 \end{cases}
\end{split}
\end{equation}
\begin{itemize}
 \item[$\mr{(iii)_B}$]If
 $P_{a\<\gamma_1,\gamma_2)}h^{1/2}_a(v)\Delta_a(v)
 P_{\<\beta_1,\beta_2)}$ and
 $P_{a\<\gamma_1,\gamma_2)}\Delta_a(v)h^{1/2}(v)
 P_{\<\beta_1,\beta_2)}$
are HS operators for a given $v\geq0$, then
$P_{a\<\gamma_1,\gamma_2)}\Delta_a(v)
 P_{\<\beta_1,\beta_2)}$ is HS for
$u\in(0,\infty)$, and
\end{itemize}
\begin{equation}\label{hs-hsen}
\begin{split}
 &\|P_{a\<\gamma_1,\gamma_2)}\Delta_a(u)
 P_{\<\beta_1,\beta_2)}\|_\HS\leq
 \frac{(\gamma_2^2+v)^{1/4}
 +(\beta_2^2+v)^{1/4}}
 {\sqrt{\gamma_1^2+u}+\sqrt{\beta_1^2+u}}\\
 &\times
 \max\big\{\|P_{a\<\gamma_1,\gamma_2)}h_a^{1/2}(v)\Delta_a(v)
 P_{\<\beta_1,\beta_2)}\|_\HS\,,
 \|P_{a\<\gamma_1,\gamma_2)}\Delta_a(v)h^{1/2}(v)
 P_{\<\beta_1,\beta_2)}\|_\HS\big\}\,.
\end{split}
\end{equation}
\begin{itemize}
\item[$\mr{(iv)_B}$] Let $\Delta^{p,q}_a(u)$ be a HS operator for
a given $u=v\geq0$. Then it is also a HS operator for
$u\in(0,\infty)$, and
\begin{align}
 &\|\Delta^{p,q}_a(u)\|_\HS<\|\Delta^{p,q}_a(v)\|_\HS&&
 \text{for}\ u>v\,,\label{hs-hsl}\\
 &\|\Delta^{p,q}_a(u)\|_\HS\leq
 \left(\frac{v}{u}\right)^{(p+q+1)/2}\|\Delta^{p,q}_a(v)\|_\HS&&
 \text{for}\ u\leq v\,,\label{hs-hss}\\
 \lim_{u\to\infty}&\|\Delta^{p,q}_a(u)\|_\HS=0\,.\label{hs-lim}
\end{align}
The operator function $(0,\infty)\ni u\mapsto \Delta^{p,q}_a(u)$
is continuously differentiable in the HS-norm sense and the formal
differentiation yields the correct result. The operator
$\Delta^{p,q}_a(0)$ is HS iff $\|\Delta^{p,q}_a(u)\|_\HS\leq\con$
for $u\in(0,\infty)$ and then $\dsp\|\Delta^{p,q}_a(0)\|_\HS=
\lim_{u\searrow0}\|\Delta^{p,q}_a(u)\|_\HS$.

The theorem remains valid upon replacement of the operator
$\Delta^{p,q}_a(u)$ with $P_{aB}\Delta^{p,q}_a(u)P_C$, for any
measurable sets $B,C$.
 \item[$\mr{(v)_B}$] Let $\Delta_a(0)$ be a HS operator. If
$\Delta_a(u)h^{1/2}(u)$ is HS for a given \mbox{$u=v\geq0$}, then
all values of the operator functions $\<0,\infty)\ni u\mapsto
\Delta_a(u)h^{1/2}(u)$ and \linebreak $\<0,\infty)\ni u\mapsto
h_a^{1/2}(u)\Delta_a(u)$ are also HS operators. Both functions are
continuous in the HS-norm on their domain and continuously
differentiable in HS-norm sense on $(0,\infty)$. Moreover,
\[
 \lim_{u\to\infty}\|h_a^{1/2}(u)\Delta_a(u)\|_\HS=0\,,\quad
 \lim_{u\to\infty}\|\Delta_a(u)h^{1/2}(u)\|_\HS=0\,.
\]

The theorem remains valid upon replacement of the operator
$\Delta_a(u)$ with $P_{aB}\Delta_a(u)P_C$, for any measurable sets
$B,C$.
\end{itemize}

\subsection{Proof of $\boldsymbol{\mr{(i)_B}}$}

For a positive real number $a$ one has
$\dsp1=\frac{1}{\pi}\int_0^\infty\frac{a}{a^2+t}\frac{dt}{\sqrt{t}}$.
Using this in the spectral representation of $h_z$ one shows that
for $u>0$
\begin{equation*}
 h(u)g(h_z)=\frac{1}{\pi}\int_u^\infty\frac{(h_z^2+u)g(h_z)}{h_z^2+t}
 \frac{dt}{\sqrt{t-u}}\,,
\end{equation*}
and the integral on the r.h.\ side converges uniformly (in norm).
With a similar representation of $f(h_{za})h_a(u)$ we have a
uniformly convergent representation
\begin{equation*}
 f(h_{za})\Delta_a(u)g(h_z)=\frac{1}{\pi}\int_u^\infty
 f(h_{za})\left[\frac{h_{za}^2+u}{h_{za}^2+t}-\frac{h_z^2+u}{h_z^2+t}\right]
 g(h_z)\frac{dt}{\sqrt{t-u}}\,.
\end{equation*}
Using the formal relation
\[
 \frac{h_{za}^2+u}{h_{za}^2+t}-\frac{h_z^2+u}{h_z^2+t}
 =(t-u)\frac{1}{h_{za}^2+t}
 \big[h_a(v)\Delta_a(v)+\Delta_a(v)h(v)\big]
 \frac{1}{h_z^2+t}\,,
\]
which becomes a correct identity when placed between $f(h_{za})$
and $g(h_z)$, one arrives at Eq.\,(\ref{hs-fg}).

\subsection{Proof of $\boldsymbol{\mr{(ii)_B}}$}

If one multiplies Eq.\,(\ref{hs-fg}) by $h^{-p}_a(u)$ from the
left and by $h^{-q}(u)$ from the right ($u>0$), one obtains a
similar identity with $\Delta_a^{p,q}$ replacing $\Delta_a$ on
both sides, and the integrand on the r.h.\ side multiplied by
$h_a^p(v)h_a^{-p}(u)$ from the left and by $h^q(v)h^{-q}(u)$ from
the right. We put $f(h_{za})=P_{a\<\gamma_1,\gamma_2)}(h_{za})$
and $g(h_z)=P_{\<\beta_1,\beta_2)}(h_z)$ in this identity, and
estimate the HS norm of the l.h.\ side. We find
\begin{multline*}
 \|P_{a\<\gamma_1,\gamma_2)}\Delta^{p,q}_a(u)
 P_{\<\beta_1,\beta_2)}\|_\HS \leq
 \|P_{a\<\gamma_1,\gamma_2)}\Delta^{p,q}_a(v)P_{\<\beta_1,\beta_2)}\|_\HS
  \\
 \times\big(\|P_{a\<\gamma_1,\gamma_2)}h_a(v)\|+\|P_{\<\beta_1,\beta_2)}h(v)\|\big)\\
 \times \|P_{a\<\gamma_1,\gamma_2)}(h_{za})h_a^{-p}(u)h_a^p(v)\|
 \|P_{\<\beta_1,\beta_2)}(h_z)h^q(v)h^{-q}(u)\|\\
 \times \frac{1}{\pi}\int_u^\infty
 \left\|\frac{P_{a\<\gamma_1,\gamma_2)}(h_{za})}{h_{za}^2+t}\right\|
 \left\|\frac{P_{\<\beta_1,\beta_2)}(h_z)}{h_z^2+t}\right\|
 \sqrt{t-u}\,dt\,,
\end{multline*}
where we have pulled the HS norm sign under the integral and used
the fact that $\|ABC\|_\HS\leq\|A\|\,\|B\|_\HS\|C\|$. The second
line in this estimate is bounded by
$\sqrt{\gamma_2^2+v}+\sqrt{\beta_2^2+v}$, the first factor in the
third line by $\max\{1,(v/u)^{p/2}\}$, the second factor in the
third line by $\max\{1,(v/u)^{q/2}\}$, and the fourth line by
\begin{equation*}
 \frac{1}{\pi}\int_u^\infty\frac{\sqrt{t-u}\,dt}{(\gamma_1^2+t)
 (\beta_1^2+t)}=\frac{1}{\sqrt{\gamma_1^2+u}+\sqrt{\beta_1^2+u}}\,,
\end{equation*}
which ends the proof of (\ref{hs-hsp}).

\subsection{Proof of $\boldsymbol{\mr{(iii)_B}}$}

We put $f(h_{za})=P_{a\<\gamma_1,\gamma_2)}(h_{za})$ and
$g(h_z)=P_{\<\beta_1,\beta_2)}(h_z)$ in the identity
(\ref{hs-fg}). Using the method of the last proof we find
\begin{multline*}
 \|P_{a\<\gamma_1,\gamma_2)}\Delta_a(u)
 P_{\<\beta_1,\beta_2)}\|_\HS \\
 \leq
 \big(\|P_{a\<\gamma_1,\gamma_2)}h_a^{1/2}(v)\|
 \|P_{a\<\gamma_1,\gamma_2)}h_a^{1/2}(v)\Delta_a(v)P_{\<\beta_1,\beta_2)}\|_\HS\\
 +\|P_{a\<\gamma_1,\gamma_2)}\Delta_a(v)h^{1/2}(v)P_{\<\beta_1,\beta_2)}\|_\HS
 \|P_{\<\beta_1,\beta_2)}h^{1/2}(v)\|\big)\\
 \times \frac{1}{\pi}\int_u^\infty
 \left\|\frac{P_{a\<\gamma_1,\gamma_2)}(h_{za})}{h_{za}^2+t}\right\|
 \left\|\frac{P_{\<\beta_1,\beta_2)}(h_z)}{h_z^2+t}\right\|
 \sqrt{t-u}\,dt\,,
\end{multline*}
which leads to the estimate (\ref{hs-hsen}).

\subsection{Proof of $\boldsymbol{\mr{(iv)_B}}$}

We set in the estimate (\ref{hs-hsp}):
\[
 \gamma_1=\sqrt{k\tau}\,,\quad
 \gamma_2=\sqrt{(k+1)\tau}\,,\quad
 \beta_1=\sqrt{l\tau}\,,\quad
 \beta_2=\sqrt{(l+1)\tau}\,,
\]
  where $k,l=0,1,\ldots$, and denote
\[
 a_{kl}(\tau,v,u)=\frac{\sqrt{k\tau+\tau+v}+\sqrt{l\tau+\tau+v}}
 {\sqrt{k\tau+u}+\sqrt{l\tau+u}}\,.
\]
Also, we introduce
$P_{a,k,\tau}=P_{a\<\sqrt{k\tau},\sqrt{k\tau+\tau})}$,
$P_{l,\tau}=P_{\<\sqrt{l\tau},\sqrt{l\tau+\tau})}$. Then from the
bound (\ref{hs-hsp}) we have
\begin{equation*}
\begin{aligned}
 &\|P_{a,k,\tau}\Delta^{p,q}_a(u)
 P_{l,\tau}\|_\HS\\
 &\hspace*{2cm}\leq
 \|P_{a,k,\tau}\Delta^{p,q}_a(v)
 P_{l,\tau}\|_\HS\,a_{kl}(\tau,v,u)\times
 \begin{cases}1&\text{for}\ u>v\,,\\
 \big(\frac{v}{u}\big)^{(p+q)/2}&\text{for}\ u\leq v\,.
 \end{cases}
\end{aligned}
\end{equation*}
If $v<u$ then we choose $\tau$ such that $v+\tau<u$, and then
$a_{kl}(\tau,v,u)<1$. If $v\geq u$, then
$a_{kl}(\tau,v,u)<[(v+\tau)/u]^{1/2}$. Moreover, there is
$\lim_{u\to\infty}a_{kl}(\tau,v,u)=0$. It is now sufficient to
observe that
\[
 \|\Delta^{p,q}_a(t)\|^2_\HS=\sum_{k,\hspace{1pt}l=0}^\infty
 \|P_{a,k,\tau}\Delta^{p,q}_a(t)P_{l,\tau}\|^2_\HS\,,
\]
to be able to conclude that Eqs.\,(\ref{hs-hsl}) and
(\ref{hs-lim}) hold. For $u\leq v$ one obtains
$\|\Delta^{p,q}_a(u)\|^2_\HS<\|\Delta^{p,q}_a(v)\|^2_\HS\,
[(v+\tau)/u](v/u)^{p+q}$ for each $\tau>0$, which leads to the
estimate (\ref{hs-hss}). Also, it should be clear that the
replacement \mbox{$\Delta^{p,q}_a(.)\to
P_{aB}\Delta^{p,q}_a(.)P_C$} poses no difficulties.

To save space we prove the remaining statements only in the case
$p=q=0$, the generalization to nonzero $p,q$ and/or added
projections $P_{aB},P_C$ is then easily obtained. We note first
that
\[
 h(t)-h(u)=\frac{t-u}{h(t)+h(u)}\,,
\]
and similarly for $h_a(.)$, are bounded operators. Thus if
$\Delta_a(t)$, $\Delta_a(u)$ are bounded then
\begin{equation*}
 \Delta_a(t)-\Delta_a(u)=-(t-u)\frac{1}{h_a(t)+h_a(u)}
 \big(\Delta_a(t)+\Delta_a(u)\big)\frac{1}{h(t)+h(u)}\,.
\end{equation*}
If on top of that $\Delta_a(t)$ and $\Delta_a(u)$ are HS, then
\[
 \|\Delta_a(t)-\Delta_a(u)\|_\HS\leq
 \frac{|t-u|}{(\sqrt{t}+\sqrt{u})^2}
 \big(\|\Delta_a(t)\|_\HS+\|\Delta_a(u)\|_\HS\big)\,.
\]
In a similar way one shows now that for $t,u>0$ there is
\begin{multline*}
 \bigg\|\frac{\Delta_a(t)-\Delta_a(u)}{t-u}+
 \frac{1}{2}h_a^{-1}(u)\Delta_a(u)h^{-1}(u)\bigg\|_\HS\\
 \leq\con\frac{|t-u|}{u^2}
 \big(\|\Delta_a(t)\|_\HS+\|\Delta_a(u)\|_\HS\big)\,.
\end{multline*}
Therefore, if the assumption is satisfied (with $v\geq0$), then
the operator function $(0,\infty)\ni u\mapsto \Delta_a(u)$ is
HS-differentiable, and its derivative is the HS-continuous
operator function $-\frac{1}{2}h_a^{-1}(u)\Delta_a(u)h^{-1}(u)$.

If $\Delta_a(0)$ is HS, then the function $\<0,\infty)\ni u\mapsto
\|\Delta_a(u)\|_\HS$ is decreasing, hence
$\dsp\|\Delta_a(0)\|_\HS\geq \lim_{t\to0}\|\Delta_a(t)\|_\HS\geq
\|\Delta_a(u)\|_\HS$ for $u\in(0,\infty)$. Conversely, if the
decreasing function $(0,\infty)\ni u\mapsto\|\Delta_a(u)\|_\HS$ is
bounded, then
\[
 \infty>\lim_{t\to0}\|\Delta_a(t)\|_\HS^2\geq\|\Delta_a(u)\|_\HS^2
 \geq\sum_{n=1}^N\|\Delta_a(u)\vp_n\|^2\,,
\]
where $\{\vp_n\}$ is any orthonormal basis. But
\begin{equation*}
  \|\Delta_a(u)-\Delta_a(0)\|=
  \bigg\|\frac{u}{h_a(u)+h_{za}}-\frac{u}{h(u)+h_z}\bigg\|
  \leq 2\sqrt{u}\,,
\end{equation*}
so taking the limit $u\searrow0$, followed by $N\to\infty$, we
have
$\dsp\lim_{t\searrow0}\|\Delta_a(t)\|_\HS\geq\|\Delta_a(0)\|_\HS$.
In conjunction with the opposite inequality obtained above this
becomes the desired equality.

\subsection{Proof of $\boldsymbol{\mr{(v)_B}}$}

If $\Delta_a(0)$ is HS, then by $\mr{(iv)_B}$ also $\Delta_a(v)$
is HS. Thus if in addition $\Delta_a(v)h^{1/2}(v)$ is HS, then by
observing the identity
\[
 \Delta_a(v)h_a(v)\Delta_a(v)=\Delta_a^3(v)+\Delta_a(v)h(v)\Delta_a(v)
\]
we learn that $h_a^{1/2}(v)\Delta_a(v)$ is also HS. The proofs of
the theorem for $\Delta_a(u)h^{1/2}(u)$ and
$h_a^{1/2}(u)\Delta_a(u)$ are similar, we take the first of these
functions. Using the estimate (\ref{hs-hsen}) we have
\begin{multline*}
 \|P_{a,k,\tau}\Delta_a(u)h^{1/2}(u)P_{l,\tau}\|^2_\HS
 \leq
 (l\tau+\tau+u)^{1/2}\|P_{a,k,\tau}\Delta_a(u)P_{l,\tau}\|^2_\HS\\
 \leq
 c^2_{kl}(\tau,v,u)
 \Big(\|P_{a,k,\tau}h_a^{1/2}(v)\Delta_a(v)P_{l,\tau}\|^2_\HS
 +\|P_{a,k,\tau}\Delta_a(v)h^{1/2}(v)P_{l,\tau}\|^2_\HS\Big)
 \,,
\end{multline*}
where
\begin{align*}
 c_{kl}(\tau,v,u)&=(l\tau+\tau+u)^{1/4}
 \frac{(k\tau+\tau+v)^{1/4}+(l\tau+\tau+v)^{1/4}}
 {\sqrt{k\tau+u}+\sqrt{l\tau+u}}\\
 &\hspace{1.5em}\leq 2(l\tau+\tau+u)^{1/4}
 \frac{(k\tau+\tau+v)^{1/4}+(l\tau+\tau+v)^{1/4}}
 {((k\tau+u)^{1/4}+(l\tau+u)^{1/4})^2}\\
 &\hspace{3em}\leq2\big(\frac{\tau+u}{u}\big)^{1/4}
 \frac{(k\tau+\tau+v)^{1/4}+(l\tau+\tau+v)^{1/4}}
 {(k\tau+u)^{1/4}+(l\tau+u)^{1/4}}\\
 &\hspace{4.5em}\leq 2\big(\frac{\tau+u}{u}\big)^{1/4}
 \max\Big\{1,\Big(\frac{\tau+v}{u}\Big)^{1/4}\Big\}
 \,.
\end{align*}
Thus $\Delta_a(u)h^{1/2}(u)$ is a HS operator for all
$u\in(0,\infty)$. Also, $\dsp\lim_{u\to\infty}c_{kl}(\tau,v,u)=0$
and $c_{kl}(\tau,v,u)<2^{5/4}$ for $u>\tau+v$, so
$\|\Delta_a(u)h^{1/2}(u)\|_\HS\to0$ for $u\to\infty$.

The HS-differentiability on $(0,\infty)$ is proved similarly as in
$\mr{(iv)_B}$. It remains to investigate $\Delta_a(0)h^{1/2}(0)$.
Using the methods of the previous proof we find
\begin{multline*}
 \Delta_a(u)h^{1/2}(u)-\Delta_a(0)h^{1/2}(0)=
 -\frac{u}{h_a(u)+h_a(0)}(\Delta_a(u)+\Delta_a(0))
 \frac{h^{1/2}(u)}{h(u)+h(0)}\\
 +\Delta_a(0)\frac{u}{[h^{1/2}(u)+h^{1/2}(0)][h(u)+h(0)]}\,.
\end{multline*}
As the r.h.\ side and the first term on the l.h.\ side are HS
operators, the operator $\Delta_a(0)h^{1/2}(0)$ is also HS.
Estimating the HS-norm of the r.h.\ side we find
\[
 \|\Delta_a(u)h^{1/2}(u)-\Delta_a(0)h^{1/2}(0)\|_\HS
 \leq u^{1/4}\big(2\|\Delta_a(0)\|_\HS+\|\Delta_a(u)\|_\HS\big)\,,
\]
which proves the missing HS-continuity at zero. Finally, it is now
easy to convince oneself that the replacement of $\Delta_a(.)$
with $P_{aB}\Delta_a(.)P_C$ poses no difficulties.

\setcounter{equation}{0}

\section{Definitions, estimates and identities}\label{pw}

In this appendix we introduce some denotations and prove some
results needed for the evaluation of the integral kernel of the
operator $h_{za}-h_z$ as defined by (\ref{mdn-mod}).

Let $F$ be a real (or complex), bounded function in
$\C^{N+2}\big(\<0,+\infty)\big)$ for some $N\geq0$, such that
$|F^{(N+2)}(p)|\leq\con\ (p+1)^{-(N+2+\gamma)}$ for some
$\gamma\in(0,1)$. Then also
\begin{equation}
 \begin{split}
 &|F^{(n)}(p)|
 \leq\con\ (p+1)^{-(n+\gamma)}\quad 1\leq n\leq N+2\,,\\
 &|F(p)-F_\infty|\leq\con\ (p+1)^{-\gamma}\,,
 \end{split}\label{pw-estF}
\end{equation}
where $F_\infty$ is a constant (limit value at infinity). We
define the following functions. For
$(q,p)\in\mR^2\setminus\{0,0\}$ we denote
\begin{equation}\label{pw-la}
 \la_F(q,p)=\frac{F(|q|)-F(|p|)}{(q-p)(q+p)}\,,
\end{equation}
so that
\begin{equation}\label{pw-las}
 \la_F(q,p)=\la_F(p,q)=\la_F(-q,p)=\la_F(q,-p)\,.
\end{equation}
We denote for $p\in\mR$
\begin{gather}
 \La_F(p)=p\int_{-\infty}^{+\infty}\la_F(q,p) \,dq
 =2p\int_0^\infty\la_F(q,p)\,dq=-\La_F(-p)\,,
 \label{pw-La}\\
 \La_{F,\ep}(p)=p\ep\!\!\sum_{k=-\infty}^{+\infty}
 \la_F(k\ep,p)=p\ep\la_F(0,p)+2p\ep\sum_{k=1}^\infty\la_F(k\ep,p)
 =-\La_{F,\ep}(-p)\,,\label{pw-Laep}\\
 \La^0_{F,\ep}(p)=p\ep\la_F(0,p)=\ep\dfrac{F(|p|)-F(0)}{p}\,,\quad
 \La^*_{F,\ep}(p)=2p\ep\sum_{k=1}^\infty\la_F(k\ep,p)\,,
 \label{pw-Laep0st}
\end{gather}
and for $p\neq p'$:
\begin{gather}
 \chi_F(p,p')=\frac{\La_F(p)-\La_F(p')}{p-p'}
 =\chi_F(p',p)=\chi_F(-p,-p')\,,
 \label{pw-chi}\\
 \chi_{F,\ep}(p,p')=\frac{\La_{F,\ep}(p)-\La_{F,\ep}(p')}{p-p'}
 =\chi_{F,\ep}(p',p)=\chi_{F,\ep}(-p,-p')\,.
 \label{pw-chiep}\\
 \chi^0_{F,\ep}(p,p')
 =\frac{\La^0_{F,\ep}(p)-\La^0_{F,\ep}(p')}{p-p'}\,,\quad
 \chi^*_{F,\ep}(p,p')
 =\frac{\La^*_{F,\ep}(p)-\La^*_{F,\ep}(p')}{p-p'}\,.
 \label{pw-chi0st}
\end{gather}

With these assumptions and denotations we have the following
results.
\begin{itemize}
\item[$\mr{(i)_C}$] The function $\la_F$ is in
$\C^{N+1}(\mR_+^2)$, all derivatives $\la_F^{(m,n)}$ for $0\leq
m+n\leq N+1$ extend to continuous functions on
$\<0,\infty)^2\setminus\{0,0\}$ and satisfy the estimates
\end{itemize}
\begin{align}
 &\text{for}\ q+p\leq1:
 &&|\la_F^{(m,n)}(q,p)|
 \leq
  \dfrac{\con}{(p+q)^{m+n+1}}\,,
  \label{pw-laest1}\\
 &\text{for}\ q+p\geq1:
 &&|\la_F^{(m,n)}(q,p)|
 \leq\con
 \begin{cases}
 \dfrac{1}{(q+1)^{m+2}(p+1)^{n+\gamma}}\,,&q\geq p\,,\\
 \dfrac{1}{(q+1)^{m+\gamma}(p+1)^{n+2}}\,,&q\leq p\,.
 \end{cases}\label{pw-laest2}
\end{align}
\begin{itemize}
\item[] The integrals $\int_0^\infty \la_F^{(m,0)}(q,p)\,dq$ are
in $\C^{N+1-m}(\mR_+)$, the differentiation may be carried out
under the integral sign, and one has the estimates:
\end{itemize}
\begin{align}
 &\text{for}\ p\leq1:\quad
 \int_0^\infty|\la_F^{(m,n)}(q,p)|\,dq\leq\con\
 \begin{cases}
 (|\log p|+1)\,,& m+n=0\,,\\
 \dfrac{1}{p^{m+n}}\,,&m+n\geq1\,,
 \end{cases}\label{pw-intla1}\\
 &\text{for}\ p\geq1:\quad
 \int_0^\infty|\la_F^{(m,n)}(q,p)|\,dq\leq\con\
 \begin{cases}
 \dfrac{1}{p^{n+1+\gamma}}\,,&m=0\,,\\
 \dfrac{1}{p^{n+2}}\,,&m\geq1\,.
 \end{cases}\label{pw-intla2}
\end{align}
\begin{itemize}
\item[$\mr{(ii)_C}$] The functions $\La_F$, $\La^*_{F,\ep}$ and
$\La^0_{F,\ep}$ are in $\C^{N+1}(\mR_+)$ and satisfy the estimates
\end{itemize}
\begin{align}
 &\text{for}\ p\leq1: \quad\begin{aligned}
 &|\La_F^{(n)}(p)|\,,\ |\La^{*(n)}_{F,\ep}(p)|\leq\con
 \begin{cases}
 p(|\log p|+1)\,,&n=0\,,\\
 (|\log p|+1)\,,&n=1\,,\\
 p^{-(n-1)}\,,&n>1\,,
 \end{cases}\\
 &|\La^{0(n)}_{F,\ep}(p)|\leq\con\,\ep\,,\end{aligned}
 \label{pw-Lab1}\\ \nonumber \\
 &\text{for}\ p\geq1:\quad|\La_F^{(n)}(p)|\,,\
 |\La^{*(n)}_{F,\ep}(p)|
 \leq \dfrac{\con}{p^{n+\gamma}}\,,\qquad
 |\La^0_{F,\ep}(p)|\leq\dfrac{\con\,\ep}{p^{n+1}}\,.
 \label{pw-Lab2}
\end{align}
\begin{itemize}
\item[] The functions $\chi_F$ and $\chi_{F,\ep}$ are in
$\C^N((\mR\setminus\{0\})^2)$ and in this domain satisfy the
estimates
\end{itemize}
\begin{align}
 &\text{for}\ \ 0<|p|\leq1\,,\ 0<|p'|\leq1:\nonumber\\
 &\hspace{40pt}|\chi_F(p,p')|\,,\ |\chi^*_{F,\ep}(p,p')|
 \leq\con\,(|\log|p||+|\log|p'||+1)\,,\label{pw-echi1}\\
 &\text{for}\ \ 0<|p|\leq1\,,\ 0<|p'|\leq1\,,\ pp'>0:\quad
 |\chi^0_{F,\ep}(p,p')|\leq\con\,\ep\,,\label{pw-echi2}\\
 &\text{for}\ \ 0<|p|\leq1\,,\ 0<|p'|\leq1\,,\ pp'<0:\quad
 |\chi^0_{F,\ep}(p,p')|\leq\dfrac{\con\,\ep}{|p|+|p'|}\,,\label{pw-echi3}\\
 &\text{for}\ \ (p,p')\in\mR^2\setminus \<-1,1\rangle^2:\nonumber\\
 &\hspace{10pt}|\chi_F(p,p')|\,,\ |\chi_{F,\ep}(p,p')|\leq
 \con
 \begin{cases}
   \dfrac{1}{(|p|+1)^\gamma(|p'|+1)}\,,&|p|\leq|p'|\,,\\
   \dfrac{1}{(|p|+1)(|p'|+1)^\gamma}\,,&|p|\geq|p'|\,.
 \end{cases}\label{pw-echi4}
\end{align}
\indent Moreover, for $|p|,|p'|\geq1$ there is
\begin{equation}\label{pw-echi5}
 |\chi^{(m,n)}_F(p,p')|\,,\ |\chi^{(m,n)}_{F,\ep}(p,p')|\leq
 \con
 \begin{cases}
   \dfrac{1}{(|p|+1)^{m+\gamma}(|p'|+1)^{n+1}}\,,&|p|\leq|p'|\,,\\
   \dfrac{1}{(|p|+1)^{m+1}(|p'|+1)^{n+\gamma}}\,,&|p|\geq|p'|\,.
 \end{cases}
\end{equation}
\begin{itemize}
\item[$\mr{(iii)_C}$] The following identity is satisfied in the
distributional sense:
\end{itemize}
\begin{equation}
 \int_{-\infty}^{+\infty}
 \frac{F(|q|)}{(p-q\pm i0)(p'-q\pm i0)}\,dq
 =\chi_F(p,p')
 \pm i\pi\frac{F(|p|)-F(|p'|)}{p-p'}\,,\label{pw-chiint}
\end{equation}
\begin{itemize}
\item[] On the l.h.\ side the distribution
 $[(p-q\pm i0)(p'-q\pm i0)]^{-1}$
is first applied to a smooth function of compact support
$f(p,p')$, and the result integrated as indicated; the r.h.\ side
is multiplied by $f(p,p')$ and integrated over $dp\,dp'$. Note
that in the integral on the l.h.\ side the signs in front of $i0$
must match.
\end{itemize}

For $p,p'\neq k\ep$ there is
\begin{gather}
 \begin{split}
 &\ep\sum_{k=-\infty}^{+\infty}
 \frac{F(|k|\ep)}{(p-k\ep)(p'-k\ep)}\\
 &\hspace{30pt}=\chi_{F,\ep}(p,p')
 -\frac{\pi}{p-p'}[F(|p|)\cot(\pi p/\ep)
 -F(|p'|)\cot(\pi p'/\ep)]\,,
 \end{split}\label{pw-chisum}\\
 \begin{split}
 &\ep\sum_{k=-\infty}^{+\infty}
 \frac{(-1)^kF(|k|\ep)}{(p-k\ep)(p'-k\ep)}\\
 &\hspace{20pt}=\chi_{F,2\ep}(p,p')-\chi_{F,\ep}(p,p')
 -\frac{\pi}{p-p'}\bigg[\frac{F(|p|)}{\sin(\pi p/\ep)}
 -\frac{F(|p'|)}{\sin(\pi p'/\ep)}\bigg]\,.
\end{split}\label{pw-chiasum}
\end{gather}
\begin{itemize}\item[]
In consequence further identities follow for all $p,p'\in\mR$:
\end{itemize}
\begin{equation}
 \begin{split}
 &2\ep\sum_{k=-\infty}^{+\infty}F(|k|\ep)
 \frac{\sin[b(p-k\ep)]\sin[b(p'-k\ep)]}{(p-k\ep)(p'-k\ep)}\\
 &\hspace{10pt}=\cos[b(p-p')]\chi_{F,\ep}(p,p')
 +\cos[b(p+p')][\chi_{F,\ep}(p,p')-\chi_{F,2\ep}(p,p')]\\
 &\hspace{120pt}+\frac{\pi\sin[b(p-p')]}{(p-p')}[F(|p|)+F(|p'|]\,,
 \end{split}\label{pw-chisin}
\end{equation}
\begin{equation}
 \begin{split}
 &2\ep\sum_{k=-\infty}^{+\infty}F(|k|\ep)(-1)^k
 \frac{\sin[b(p+k\ep)]\sin[b(p'-k\ep)]}{(p+k\ep)(p'-k\ep)}\\
 &=-\cos[b(p-p')]\chi_{F,\ep}(p,-p')
 -\cos[b(p+p')][\chi_{F,\ep}(p,-p')-\chi_{F,2\ep}(p,-p')]\\
 &\hspace{120pt}-\frac{\pi\sin[b(p-p')]}{(p+p')}[F(|p|)-F(|p'|]\,,
 \end{split}\label{pw-chiasin}
\end{equation}

\subsection{Proof of $\boldsymbol{\mr{(i)_C}}$}

For $q,p>0$ one has
\begin{equation}\label{pw-difr}
 \frac{F(q)-F(p)}{q-p}=\int_0^1 F^{(1)}(qt+p(1-t))\,dt\,,
\end{equation}
so by (\ref{pw-estF}) the differential properties of
$\la_F^{(m,n)}$ are satisfied. For $p+q\leq1$ the derivatives of
(\ref{pw-difr}) are bounded, so the estimate (\ref{pw-laest1}) is
also true. To prove the estimate (\ref{pw-laest2}) it is
sufficient to assume that $q+p\geq1$ and $q\geq p$ (due to the
symmetry of $\la_F$), which implies $q+p\geq(1+q)/2$ and
\begin{equation}\label{pw-dqpp}
 \left|\p_q^k\p_p^l\frac{1}{q+p}\right|\leq\frac{\con}{(q+1)^{k+l+1}}\,.
\end{equation}
 We consider two cases $q\leq 3p$ and $q>3p$ separately. In the
first of these regions one has $p\geq q/3\geq(q+p)/6\geq(q+1)/12$
and
\begin{equation}
\begin{split}
 \left|\p_q^{r}\p_p^{s}\frac{F(q)-F(p)}{q-p}\right|\leq
 &\int_0^1|F^{(r+s+1)}(qt+p(1-t))|\,dt\\
 \leq\con\ &\int_0^1\frac{dt}{(p+(q-p)t)^{r+s+1+\gamma}}\leq
 \frac{\con}{(q+1)^{r+s+1+\gamma}}\,,
\end{split}
\end{equation}
so $|\la^{(m,n)}(q,p)|\leq\con\ (q+1)^{-(m+n+2+\gamma)}$ in this
region, which complies with the estimate (\ref{pw-laest2}). In the
second region one has $q-p=(q+p+q-3p)/2\geq(q+1)/4$, so taking
into account Eq.\,(\ref{pw-dqpp}) one has in that region:
\begin{equation}\label{pw-dqp}
 \left|\p_q^k\p_p^l\frac{1}{(q-p)(q+p)}\right|
 \leq\frac{\con}{(q+1)^{k+l+2}}\,.
\end{equation}
Taking also into account the bounds (\ref{pw-estF}), and similar
ones with the argument $p$ replaced by $q$, one confirms the
estimate (\ref{pw-laest2}).

The statements following estimates (\ref{pw-laest1}) and
(\ref{pw-laest2}) are their simple consequences. The next two
estimates (\ref{pw-intla1}) and (\ref{pw-intla2})) are obtained
from the preceding two by elementary integration.

\subsection{Proof of $\boldsymbol{\mr{(ii)_C}}$}

For $\La_F$ the differentiability properties and the estimates
(\ref{pw-Lab1}) and (\ref{pw-Lab2}) are simple applications of the
properties stated in the last sentence of $\mr{(i)_C}$. To prove
the same facts for $\La_{F,\ep}$ one has to estimate the sums by
integrals, which can be done rather easily by observing that for
each $p>0$ the r.h.\ sides of the bounds (\ref{pw-laest1}) and
(\ref{pw-laest2}) can be glued together into a continuous
decreasing function of $q$.\linebreak For $\La_{F,\ep}^0$ the
estimates (\ref{pw-Lab1}) and (\ref{pw-Lab2}) follow from
(\ref{pw-difr}) and (\ref{pw-laest2}) respectively. The estimates
(\ref{pw-echi5}) in the enlarged region $|p|,|p'|\geq1/2$ follow
from the bounds (\ref{pw-Lab2}) (which stay valid for $|p|>1/2)$
with the use of the method similar to that applied in the proof of
$\mr{(i)_C}$. The estimates (\ref{pw-echi4}) are then easily
extended to the whole region of application with the use of bounds
(\ref{pw-Lab1}), (\ref{pw-Lab2}) (with $n=0$). The remaining
estimates also follow from these bounds by cutting the square
$\<-1,1\rangle^2$ into four squares according to the signs of $p$
and $p'$.

\subsection{Proof of $\boldsymbol{\mr{(iii)_C}}$}

To prove (\ref{pw-chiint}) we first note that for complex $z,z'$
with $\Im z,\Im z'\neq0$ there is
\begin{equation*}
    z\int_{-\infty}^{+\infty}\frac{F(|q|)\,dq}{(q-z)(q+z)}
    -z'\int_{-\infty}^{+\infty}\frac{F(|q|)\,dq}{(q-z')(q+z')}
    =(z-z')\int_{-\infty}^{+\infty}\frac{F(|q|)\,dq}{(q-z)(q-z')}
\end{equation*}
(use evenness of the integrand on the r.h.\ side). Let
$1>\alpha>\alpha'>0$. Setting $z=p\pm i\alpha$, $z'=p'\pm
i\alpha'$, denoting
\begin{equation*}
    \La_F^{\pm\alpha}(p)=(p\pm i\alpha)\int_{-\infty}^{+\infty}
    \frac{F(|q|)-F(|p|)}{(q-(p\pm i\alpha))(q+p\pm i\alpha)}
    \,dq\,,
\end{equation*}
and using the identity
\[
 \int_{-\infty}^{+\infty}
 \frac{dq}{(q-(p\pm i\alpha))(q+p\pm i\alpha)}
 =\pm\frac{\pi i}{p\pm i\alpha}
\]
we get
\begin{equation}\label{pw-chiintal}
 \int_{-\infty}^{+\infty}
 \frac{F(|q|)\,dq}{(p-q\pm i\alpha)(p'-q\pm i\alpha')}
 =\frac{\La_F^{\pm\alpha}(p)-\La_F^{\pm\alpha'}(p')}
 {p-p'\pm i(\alpha-\alpha')}
 \pm i\pi\frac{F(|p|)-F(|p'|)}{p-p'\pm i(\alpha-\alpha')}
\end{equation}
We multiply this identity by a Schwartz function $f(p,p')$,
integrate, and take successive limits $\alpha'\to0$, $\alpha\to0$.
Now, for the l.h.\ side we observe that the order of integration
may be changed and the result represented as an integral over $q$
of $F(|q|)$ multiplied by
\[
 \int\frac{f(p,p')}{(p-q\pm i\alpha)(p'-q\pm
 i\alpha')}\,dp\,dp'\,.
\]
By the well known techniques one shows that the consecutive limit
exists and, moreover, the above expression is bounded by
$\con\,(q^2+1)^{-1}$. This is sufficient to conclude that l.h.\
side of Eq.\,(\ref{pw-chiint}) is obtained in the limit. For the
r.h.\ side of (\ref{pw-chiintal}) one easily notes that its
consecutive point-wise limit yields for $p\neq p'$ the function on
the r.h.\ side of (\ref{pw-chiint}). Thus to complete the proof
one only needs to show that in the limiting process the r.h.\ side
of (\ref{pw-chiintal}) stays bounded by a function defining a
distribution. This is immediate for the second term, as
\[
 \bigg|\frac{F(|p|)-F(|p'|)}{p-p'\pm i(\alpha-\alpha')}\bigg|
 \leq|p+p'|\,\big|\la_F(p,p')\big|\,.
\]
Also, the property is rather obvious for the limiting process
$\alpha'\to0$ in the first term (note that
$|\La_F^{\pm\alpha}(p)|\leq(|p|+1)\int|\la_F(q,p)|dq$). Thus we
are left with the function
\begin{equation*}
 \frac{\La_F^{\pm\alpha}(p)-\La_F(p')}
 {p-p'\pm i\alpha}
 =\frac{\La_F^{\pm\alpha}(p)-\La_F(p)}
 {p-p'\pm i\alpha}+
 \frac{\La_F(p)-\La_F(p')}
 {p-p'\pm i\alpha}\,.
\end{equation*}
The second term on the r.h.\ side of this equality is bounded by
$|\chi_F(p,p')|$, which is sufficient for our purpose, while for
the first one it is sufficient to estimate the function
\begin{equation}\label{pw-lap}
 \begin{split}
  \frac{\La_F^{\pm\alpha}(p)-\La_F(p)}{\pm i\alpha}
  &=\int\frac{[F(|q|)-F(|p|)]\ dq}{(q-(p\pm i\alpha))(q+p\pm i\alpha)}
  \pm i\pi p\frac{2p\pm i\alpha}{p\pm i\alpha}\la_F(p,p)\\
  &+p(2p\pm i\alpha)\int
  \frac{\la_F(q,p)-\la_F(p,p)}{(q-(p\pm i\alpha))(q+p\pm i\alpha)}
  \,dq\,.
 \end{split}
\end{equation}
The first two terms on the r.h.\ side of this identity are bounded
respectively by $\int|\la_F(q,p)|dq$ and $2\pi|p\la_F(p,p)|$,
which fulfills our demands. For the third term we assume that
$p>0$ (the case $p<0$ needs only obvious modifications), represent
the integral as twice the integral over $\<0,\infty)$, and
consider the integration sets $\<0,p+1)$ and $\<p+1,\infty)$
separately. The term containing integration over the second set is
bounded by
\[
 (p+1)\log(1+2p)\Big(|\la_F(p,p)|+\max_{q\in\<p+1,\infty)}|\la_F(q,p)|
 \Big)\leq\con\frac{\log(1+2p)}{(p+1)^{1+\gamma}}\,.
\]
The remaining part is bounded by
\[
 2p\int_0^{p+1}\bigg|\frac{\la_F(q,p)-\la_F(p,p)}{q-p}\bigg|\,dq
 \leq 2p\int_0^{p+1}
 \frac{1}{|q-p|}\int_{r_<}^{r_>}\big|\la_F^{(1,0)}(s,p)\big|\,ds\
 dq\,,
\]
where $r_<=\min\{q,p\}$, $r_>=\max\{q,p\}$. For $p>1$ there is
$|\la_F^{(1,0)}(s,p)|\leq\con\, p^{-2}$ (cf.\ (\ref{pw-laest2})),
which gives a sufficient estimate in this region. For $p<1$ and
$q<p+1$ one has $|\la_F(s,p)|\leq\con\,(s+p)^{-2}$ (by
(\ref{pw-laest1}), which may be used in this extended region).
Thus the r.h.\ side is bounded by $\con\log[(2p+1)/p]$, which ends
the proof of (\ref{pw-chiint}).

To prove (\ref{pw-chisum}) we first note that for $p,p'\neq k\ep$,
there is
\[
 (p-p')\,\ep\!\sum_{k\in\mZ}
 \frac{F(|k|\ep)}{(p-k\ep)(p'-k\ep)}
 =p\,\ep\!\sum_{k\in\mZ}\frac{F(|k|\ep)}{(k\ep-p)(k\ep+p)}
 -p'\ep\!\sum_{k\in\mZ}\frac{F(|k|\ep)}{(k\ep-p')(k\ep+p')}\,.
\]
But
\[
 p\,\ep\!\sum_{k\in\mZ}\frac{F(|k|\ep)}{(k\ep-p)(k\ep+p)}
 =\La_{F,\ep}(p)-F(|p|)\pi\cot(\pi p/\ep)\,,
\]
where we used the following identity
\[
 x\sum_{k\in\mZ}\frac{1}{x^2-k^2}=\pi\cot(\pi x)\,,\quad
 (x\notin\mZ)\,
\]
(cf.\,\cite{gr}, formula 1.421(3)). This ends the proof of
(\ref{pw-chisum}).

Formula (\ref{pw-chiasum}) follows easily from the preceding one
if one observes that its l.h.\ side may be written as
$2\ep\sum\dfrac{F(|k|2\ep)}{(p-k2\ep)(p'-k2\ep)}-
\ep\sum\dfrac{F(|k|\ep)}{(p-k\ep)(p'-k\ep)}$. Finally, to prove
(\ref{pw-chisin}) and (\ref{pw-chiasin}) one notes that
\[
 2\sin[b(p-k\ep)]\sin[b(p'-k\ep)]=\cos[b(p-p')]-(-1)^k\cos[b(p+p')]
\]
and uses (\ref{pw-chisum}) and (\ref{pw-chiasum}).

\setcounter{equation}{0}

\section{Some identities for the logarithmic derivative of the
Gamma function $\boldsymbol{\psi}$}\label{apsi}

Two particular textbook representations of the function $\psi$ are
of importance for us:
\begin{gather}
  \psi(z)=-\frac{1}{z}+\lim_{N\to\infty}
  \bigg\{\log N-\sum_{k=1}^N\frac{1}{k+z}\bigg\}\,,\label{psi-rep1}\\
  \psi(z)-\log z+\frac{1}{z}
  =\int_0^\infty v(s)e^{-zs}\,ds\,,\quad \Re z>0\,,
  \label{psi-rep2}
\end{gather}
where
\begin{equation}\label{psi-v}
    v(s)=\frac{1}{s}-\frac{1}{e^s-1}\,.
\end{equation}
The function $v$ is analytical in the complex plane outside the
points $z=2k\pi i$, $k\in\mathbb{Z}\setminus\{0\}$, where it has
poles with principal values $-1$. Moreover,
\begin{equation}\label{psi-v2}
    v^{(k)}(0)=-\frac{B_{k+1}}{k+1}\,,\qquad
    |v^{(k)}(s)|\leq\frac{c_k}{(|s|+1)^{k+1}}\,,\quad s\in\mR\,,\
    k=0,1,\ldots\,.
\end{equation}
Using these properties in the representation (\ref{psi-rep2}) one
finds by induction for $m\in\mathbb{N}$ and $\Re z>0$ the
expansion:
\begin{equation}\label{psi-exp}
    \psi(z)-\log z
    =-\frac{1}{2z}-\sum_{k=1}^m\frac{B_{2k}}{2k}\frac{1}{z^{2k}}
    +\frac{1}{z^{2m}}\int_0^\infty v^{(2m)}(s)e^{-zs}\,ds
\end{equation}
(remember that $B_{2k+1}=0$ for $k\in\mathbb{N}$). We denote
\begin{equation}\label{psi-w}
    w_{2m}(z)=z^{2m}\bigg\{\psi(z)-\log z
    +\frac{1}{2z}
    +\sum_{k=1}^m\frac{B_{2k}}{2k}\frac{1}{z^{2k}}\bigg\}\,.
\end{equation}
Then for $\la\geq0$, $m=1,2,\ldots$, one has the identity
\begin{equation}\label{psi-F}
    \int_0^\infty\cos(2\pi\lambda t)w_{2m}(t)\,dt
    =(-1)^{m-1}\frac{(2m)!}{2(2\pi)^{2m}}
    \sum_{k=1}^\infty\frac{1}{(k+\lambda)^{2m+1}}\,.
\end{equation}
In particular,
\begin{equation}\label{psi-F2}
    \int_0^\infty\cos(2\pi\lambda t)w_2(t)\,dt
    =\frac{1}{4\pi^2}
    \sum_{k=1}^\infty\frac{1}{(k+\lambda)^3}\,,
\end{equation}
and by the application of the operator $\la\p_\la$:
\begin{equation}\label{psi-F2prim}
    \int_0^\infty\cos(2\pi\lambda t)tw^{(1)}_2(t)\,dt
    =\frac{1}{4\pi^2}
    \bigg\{3\la\sum_{k=1}^\infty\frac{1}{(k+\la)^4}
    -\sum_{k=1}^\infty\frac{1}{(k+\lambda)^3}\bigg\}\,.
\end{equation}

To prove Eq.\,(\ref{psi-F}) one notes that
$v^{(2m)}(-s)=-v^{(2m)}(s)$ and uses Eq.\,(\ref{psi-exp}) to
obtain for $\lambda>0$
\begin{equation}\label{psi-pr}
    \int_0^\infty\cos(2\pi\lambda t)w_{2m}(t)\,dt
    =\frac{1}{2}\int_{-\infty}^{+\infty}\frac{v^{(2m)}(s)}{s+i2\pi\la}\,ds
    =\frac{(2m)!}{2}\int_{-\infty}^{+\infty}\frac{v(s)}{(s+i2\pi\la)^{2m+1}}\,ds\,.
\end{equation}
The proof is now completed for $\lambda>0$ by integration in the
complex $s$-plane along the rectangular contour with vertices at
points $\pm\xi$, $\pm\xi+(2k+1)\pi i$ followed by the successive
limits $\xi\to\infty$ and $k\to\infty$. Finally, one uses the
continuity in $\lambda$ of both sides of (\ref{psi-F}) to extend
the formula to $\lambda=0$.

\setcounter{equation}{0}

\end{document}